\documentclass[aps,pra,twocolumn,amsmath,amssymb, superscriptaddress,tightenlines]{revtex4}
\usepackage{graphicx}
\usepackage{epsfig}
\usepackage{dcolumn}
\usepackage{bm}
\usepackage{amssymb}
\usepackage{bm}
\usepackage{amsmath, bm}
\usepackage{upgreek}
\usepackage[colorlinks,citecolor=blue,linkcolor=blue,hyperindex]{hyperref}

\begin{document}

\title{Partial equilibration of integer and fractional edge channels in the thermal quantum Hall effect}
\author{Ken K. W. Ma and D. E. Feldman}
\affiliation{Department of Physics, Brown University, Providence, Rhode Island 02912, USA}
\date{\today}
 

  
\begin{abstract}
Since the charged mode is much faster than the neutral modes on quantum Hall edges at large filling factors, the edge may remain out of equilibrium in thermal conductance experiments. This sheds light on the observed imperfect quantization of the thermal Hall conductance at $\nu=8/3$ and can increase the observed thermal conductance by two quanta at $\nu=8/5$. Under certain unlikely but not impossible assumptions, this might also reconcile the observed thermal conductance at $\nu=5/2$ with not only the PH-Pfaffian order but also the anti-Pfaffian order.
\end{abstract}

\maketitle

\section{Introduction}

A great majority of the known fractional quantum Hall plateaus can be understood as integer plateaus of composite fermions \cite{jain-book}. The picture of weakly interacting composite fermions predicts compressible phases at even-denominator filling factors, as observed indeed at $\nu=1/2$. A theory of incompressible states at half-integer filling factors involves an additional idea that composite fermions form Cooper pairs \cite{x41-RG}. This idea leads to much interesting physics, including non-Abelian statistics \cite{2-MR}. It also explains why the even-denominator quantum Hall effect (QHE) has been a challenging problem for decades: Different pairing channels produce numerous topological orders at the same filling 
factor \cite{foot0}. In a marked contrast, it is hard to find a viable alternative to the picture of the $\nu=1/3$ QHE plateau as a $\nu=1$ state of composite fermions.
The nature of the $\nu=5/2$ QHE liquid remains controversial. 

Numerical work on $\nu=5/2$ has much history, and different topological orders were seen as leading candidates at different times \cite{HR-1988,myg,x1-Morf,new-num,Rezayi-2017}. Most recently, a preponderance of numerical evidence \cite{Rezayi-2017} has been pointing out at the anti-Pfaffian topological order \cite{x3-LRH,x4-LRNF} in translationally invariant systems \cite{BoYang}. Experimental evidence appears to consistently point \cite{x12-ZF} towards a closely related but distinct PH-Pfaffian topological 
order \cite{x4-LRNF,x18-Son,x18a-Ashwin,x18b-Bonderson}. 
Both numerics and experiment have limitations. Indeed, the existing numerical work always neglects some important features of realistic samples, such as disorder \cite{disorder-num}, and treats Landau level mixing (LLM) in a highly approximate way. At the same time, the experimental evidence of topological order is rather indirect.

The most direct observation in favor of the PH-Pfaffian hypothesis has come from a thermal conductance experiment \cite{x19-Nature-2018}. The thermal conductance of an equilibrium QHE edge is quantized at $KT=n\kappa_0 T$, where one thermal conductance quantum $\kappa_0 T=\pi^2k_B^2 T/3h$, and a universal prefactor $n$ 
depends on the topological order \cite{x22-KF-eq,Capelli}. 
The measured thermal conductance $KT\approx 2.5\kappa_0 T$ at $\nu=5/2$ and the bath temperature $T_0\sim 20~$mK is consistent with the PH-Pfaffian order \cite{x18-Son,x12-ZF}. The thermal conductance of an anti-Pfaffian liquid \cite{x3-LRH,x4-LRNF} is $1.5\kappa_0 T$.

The interpretation of the $\nu=5/2$ data crucially depends on the agreement between theory and experiment at other filling factors. Good agreement has been achieved at all integer filling factors \cite{x21-Science-2013,x24-Nature-2017} and at the fractional filling factors $\nu=1/3, 4/7, 3/5,$ and $7/3$ \cite{x19-Nature-2018,x24-Nature-2017}. The $\nu=2/3$ data \cite{x24-Nature-2017} appeared to conflict with the theory \cite{x22-KF-eq}. The apparent conflict was explained  by unusually slow edge equilibration at $\nu=2/3$ due to the unusual edge structure with the same numbers of the modes, propagating in the two opposite directions (see the Methods section in Ref. \onlinecite{x24-Nature-2017}). Thus, the only filling factor with a large unexplained discrepancy of the theory with the experiment is $\nu=8/3$. Yet, this filling factor is close to $\nu=5/2\approx 8/3$, and hence the observed discrepancy puts a question mark over the meaning of the $\nu=5/2$ data. This motivates the main goal of the present paper to understand the imperfect quantization of the $\nu=8/3$ thermal conductance.

Our main idea follows from the observation that the charged mode with the conductance $8e^2/3h$ is much faster than the neutral modes on an etched edge. A high speed means a low density of states for the excitations of the charged mode. This suppresses heat flow between the charged mode and the rest of the system and thus effectively decouples the charged mode. We also argue that the spin mode decouples from the rest of the edge channels. The charged mode, the spin mode, and the two remaining strongly interacting edge channels form three decoupled systems that carry heat in parallel. The first two make a universal contribution $2\kappa_0 T$ to the total heat conductance. The third one behaves like the $\nu=2/3$ edge and adds a non-universal contribution to the heat conductance. We thus reconcile the observed $\nu=8/3$ thermal transport with the theory, and hence find that one can trust the experimental data at $\nu=5/2\approx 8/3$.

The second motivation of this paper is an attempt to reconcile the observed thermal conductance at $\nu=5/2$ with a possibility of the numerically supported anti-Pfaffian order. This may seem unpromising. Indeed, several recent studies show how disorder \cite{x43-d1,x44-d2,x45-d3} and LLM \cite{Milovanovic} can stabilize a PH-Pfaffian liquid in a realistic sample. It is thus tempting to ascribe the tension between numerics and experiment to the simplified structure of the numerical Hamiltonians. The only attempt \cite{x25-Simon} so far to interpret the data in terms of the anti-Pfaffian state faces difficulties (see Ref. \onlinecite{comment} for a detailed discussion of those difficulties and the reasons why the challenges of the present approach to $\nu=5/2$ are subtler). In this paper, we develop a different physical picture that builds on the high velocity of the charged mode with the conductance $5e^2/2h$. We find that the PH-Pfaffian order explains the data better, but do not  completely rule out a possibility of an anti-Pfaffian liquid, given the limited amount of data and its limited accuracy. Due to the importance of the $5/2$ problem, it is crucial to explore all possible interpretations. As a part of this work, we address what experiments can fully settle the issue of the topological order at $\nu=5/2$.

Our approach follows the $\nu=8/3$ case. Three subsystems conduct heat in parallel: the charged mode, the spin mode, and a system of four neutral modes. The charged mode and the spin mode contribute one quantum of thermal conductance each. The four remaining modes on an anti-Pfaffian edge include one Bose mode and three Majorana modes, propagating in the opposite direction. Under the assumption that those four modes are in equilibrium with each other, they contribute the thermal conductance $(3\times 1/2 -1)\kappa_0 T=\kappa_0 T/2$. The total thermal conductance $2.5\kappa_0 T$ is then consistent with the experiment.

The key challenge to the above picture is rapid growth of the observed $K=[{\rm thermal~conductance}]/T$ at low bath temperatures. This has a natural explanation in the PH-Pfaffian state (see the Methods section in Ref. \onlinecite{x19-Nature-2018}). We have to invoke fine-tuning to reconcile the observed temperature dependence of the thermal conductance with the anti-Pfaffian order. This suggests that the PH-Pfaffian picture works better than the anti-Pfaffian hypothesis.

Our physical picture allows more definite conclusions about several other filling factors in the upper spin branch of the first Landau level. We predict that at some filling factors the experimental technique \cite{x19-Nature-2018} can yield the thermal conductance value that exceeds by $2\kappa_0 T$ the equilibrium thermal conductance of an infinite sample.
The effect is most dramatic at $\nu=8/5$. An equilibrium edge is expected to have zero thermal conductance at $\nu=8/5$. We instead predict $K_{8/5}T=2\kappa_0 T$. This shows that the success of the thermal conductance experiments \cite{x19-Nature-2018,x24-Nature-2017} is to a large extent a matter of luck with the choice of the filling factors to study.

The article is organized as follows. We start with a brief review of topological orders in the second Landau level and in the upper spin branch of the lowest Landau level. 
Next, we discuss edge physics at $\nu=8/3$ in Section~\ref{sec:edge_8/3}. The results are used to understand heat exchange between 
the charged mode and the neutral modes in Section~\ref{sec:length}. 
We extend that analysis to $\nu=8/5$ in Section~\ref{sec:8/5_prediction}. Section~\ref{sec:thermal_8/3} explains poor quantization of the thermal conductance at $\nu=8/3$. We attempt to reconcile the observed $KT\approx 2.5\kappa_0 T$ with the anti-Pfaffian order at $\nu=5/2$ in Section~\ref{sec:5/2}. The final section summarizes our results. Six Appendices address technical details and subtle points. Appendix \ref{app:voltage} addresses voltage equilibration between edge channels. Appendix \ref{app:losses} considers energy losses from the edge to the bulk. Appendix \ref{app:pheno-model} develops a phenomenological model of heat transport at $\nu=8/3$.  Small corrections to thermal conductance are addressed in 
Appendix \ref{app:correction}. A subtle role of the $RC$-time is discussed in Appendix \ref{app:RC}. Possible breakdown of bulk-edge correspondence is the subject of 
Appendix \ref{app:bulk-edge}.

\section{Review of topological orders}
\label{sec:review}

Three fractional filling factors $7/3,$ $5/2,$ and $8/3$ from the second Landau level were observed in the thermal conductance experiment with the sample \cite{x19-Nature-2018}. We start with a brief review of those filling factors. We then discuss the upper spin branch of the lowest Landau level with the emphasis on $\nu=8/5$. The edge structures at the relevant filling factors are summarized in Table \ref{tab:1}.

\subsection{$\nu=7/3$}
\label{subsec:7/3}

The topological order is believed to be the same as in the Laughlin state at $\nu=1/3$. This is supported by the observed thermal conductance \cite{x19-Nature-2018}, the absence of an upstream mode on the edge \cite{8-3-2011}, and the observed \cite{8-3-2011} quasiparticle charge $e/3$. This is also consistent with the observed $100\%$ polarization of the electrons in the second Landau level \cite{Tiemann-NMR,Pan-spin}.

The filled first Landau level gives rise to two integer edge channels. We call their propagation direction {\it downstream}. The opposite direction is {\it upstream}. There is also one downstream fractional edge channel. Its physics is the same as in the $1/3$-state \cite{wen-book}. Each edge channel carries one quantum of thermal conductance. Thus, the theoretical thermal conductance $K_{7/3}T=3\kappa_0 T$.

\subsection{$\nu=8/3$}

The topological order of this state is believed to be the same as at $\nu=2/3$. It is the particle-hole conjugate of the Laughlin order at $\nu=1/3$. This picture receives support from the same experiments as at $\nu=7/3$: the measured quasiparticle charge is $e/3$ \cite{8-3-2011}, 
and a topologically-protected upstream neutral mode was detected on the edge \cite{8-3-2011}. 
As we argue below, the observed thermal conductance \cite{x19-Nature-2018} is also compatible with the same topological order as at $\nu=2/3$. 

In contrast to the $\nu=7/3$ liquid, the 
$\nu=8/3$ liquid is believed to have different spin polarizations at different charge densities \cite{Pan-spin}. The system is fully polarized \cite{Tiemann-NMR,Pan-spin,Stern-NMR} at the densities, employed in Ref. \onlinecite{x19-Nature-2018}. Note that an optical experiment \cite{Stern-optical} was interpreted as showing zero spin polarization at $\nu=8/3$ at charge densities, relevant for Ref. \onlinecite{x19-Nature-2018}. This interpretation was later questioned \cite{Stern-NMR,Rhone-2011}. A subsequent optical experiment
\cite{Wurstbauer2013} suggests that Ref. \onlinecite{Stern-optical} underestimates spin polarization.

Like in all other states of the second Landau level, the edge carries two integer downstream modes. There are also two fractional modes. Their simplest picture can be obtained by the particle-hole conjugation of the edge structure at $\nu=1/3$. This way one finds two charged modes: a downstream mode with the conductance $e^2/h$ and an upstream mode of the conductance $e^2/3h$. Electron tunneling and Coulomb interaction greatly modify this picture, as discussed in the next section.

Each of the four edge modes carries one quantum of thermal conductance $\kappa_0 T$. Three modes transport heat downstream and one mode transports heat upstream. In thermal equilibrium, all modes have the same temperature, and  the combined thermal conductance $K_{8/3}T=(3-1)\kappa_0 T=2\kappa_0 T$.

\subsection{$\nu=5/2$}

Numerous Abelian and non-Abelian topological orders have been proposed at $\nu=5/2$. 
 See Refs. \onlinecite{foot0,x19-Nature-2018} for a review. Experiment shows evidence \cite{x40-composite,xnew-composite} of composite fermions at $\nu=5/2$. In agreement with this evidence, all leading candidate topological orders at that filling factor can be understood in terms of Cooper pairs of composite fermions \cite{x41-RG}. 

\begin{figure}[htb]
\includegraphics[width=3.0 in]{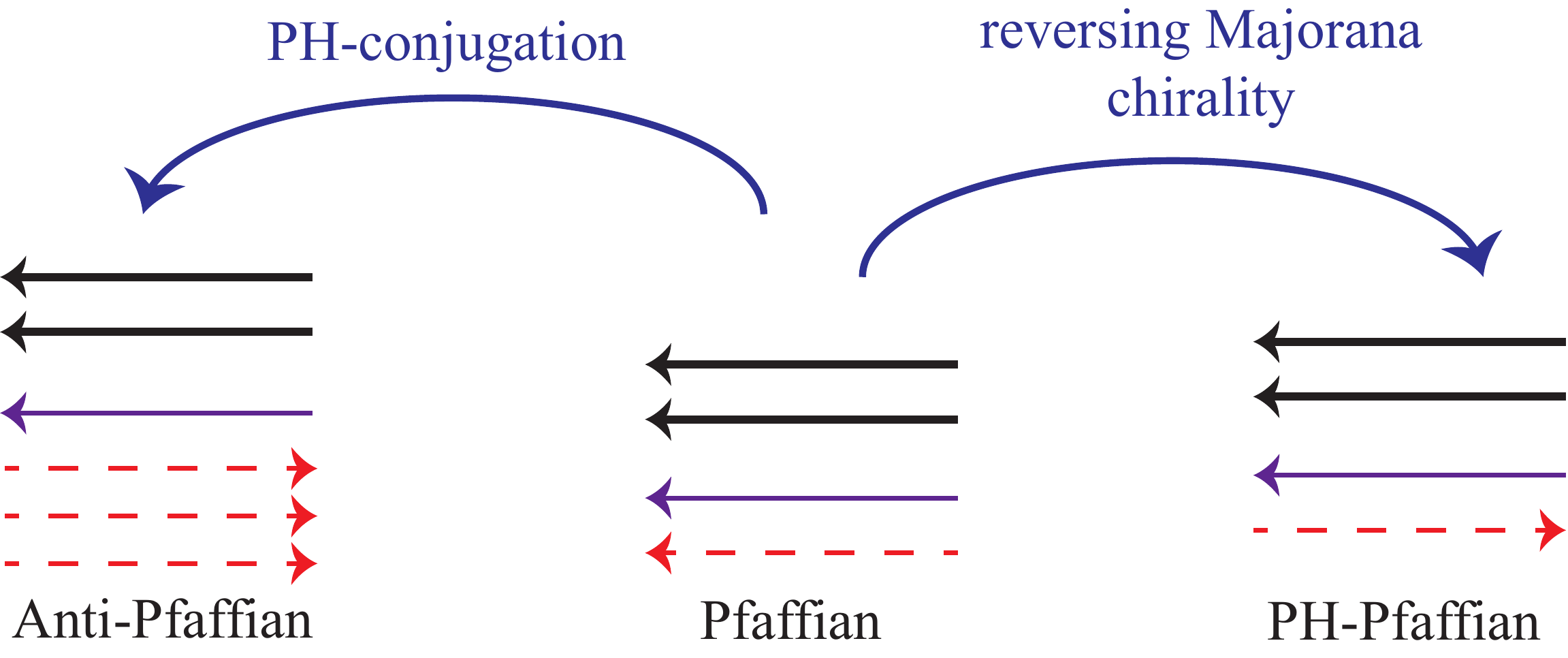}
\caption{(color online) Edge modes for the three leading non-Abelian contenders for the $\nu=5/2$ QHE state: Pfaffian, anti-Pfaffian, and PH-Pfaffian. The two thick black lines depict an integer charged mode with the conductance $2e^2/h$ and an integer spin mode. The thin purple line depicts a fractional charged mode with the conductance $e^2/2h$. The red dashed lines depict neutral Majorana fermions.}
\label{fig:A}
\end{figure}

Our main focus will be on the PH-Pfaffian and anti-Pfaffian orders. Both are close relatives of the Pfaffian liquid \cite{2-MR}, and we start with its brief review. 
We will be mostly interested in the edge structure for these states. The edge structures are illustrated in Fig. \ref{fig:A}.
 
The simplest wave function with the Pfaffian order
\begin{align}
\label{dima-new-1}
&\Psi_{\rm Pf}(\{z_i\})
\nonumber\\
=&~{\rm Pf}\{\frac{1}{z_i-z_j}\}\Pi(z_i-z_j)^2\exp(-\sum |z_i|^2/4l_B^2),
\end{align}
where $z_k=x_k+iy_k$ are the positions of the electrons, and $l_B$ is the magnetic length. The Pfaffian factor encodes the $p+ip$ pairing of composite fermions \cite{x41-RG}. In addition to two Bose integer modes, the edge theory contains a downstream fractional charged Bose mode of the conductance $e^2/2h$ and a neutral downstream Majorana.   Each of the three Bose modes carries one quantum of thermal conductance. A Majorana mode carries half a quantum \cite{x19-Nature-2018}. Thus, the total thermal conductance $K_{\rm Pf}T=3.5\kappa_0 T$.

The anti-Pfaffian state is the particle-hole conjugate of the Pfaffian liquid \cite{x3-LRH,x4-LRNF}. This corresponds to $f$-pairing of composite fermions. The edge structure involves two integer Bose modes, one more downstream Bose mode of the conductance $e^2/h$, and two upstream modes corresponding to two fractional modes in the Pfaffian state. One is a charged boson and the other is a neutral Majorana. Disorder is inevitably present on an edge \cite{comment} and is known to greatly modify this picture \cite{x3-LRH,x4-LRNF}. The resulting low-energy effective theory on the edge includes two integer channels, a downstream fractional channel of the conductance $e^2/2h$, and three upstream Majorana modes of equal velocity. Both pictures are equivalent for the calculation of the equilibrium heat conductance. If all modes are in equilibrium with the same temperature, the thermal conductance 
$K_{\rm aPf}T=(3-3\times 0.5)\kappa_0 T=1.5\kappa_0 T$.

The PH-Pfaffian order is intermediate between the Pfaffian and anti-Pfaffian orders \cite{x18-Son}. The simplest wave function \cite{x12-ZF} is obtained by the complex conjugation of the Pfaffian factor in  Eq. (\ref{dima-new-1}). This corresponds to $p-ip$ pairing of composite fermions. The edge structure \cite{x12-ZF} differs from the Pfaffian liquid by the opposite direction of the Majorana mode. Hence, the equilibrium thermal conductance $K_{\rm PH}T=(3-0.5)\kappa_0 T=2.5\kappa_0 T$ in agreement 
with the experiment \cite{x19-Nature-2018}. This value is the average of $K_{\rm Pf}T$ and $K_{\rm aPf}T$, 
and it was proposed that the macroscopic PH-Pfaffian order would emerge in a mixture of microscopic domains with Pfaffian 
and anti-Pfaffian liquids \cite{x43-d1,x44-d2,x45-d3}. 
A similar relation may exist between the PH-Pfaffian order and other pairs of particle-hole conjugate orders. 
For example, the $SU(2)_2$ and 
anti-$SU(2)_2$ orders have the thermal conductances \cite{x19-Nature-2018} of $4.5\kappa_0 T$ and $0.5\kappa_0 T$. Their mixture might also lead to the macroscopic PH-Pfaffian order.

All five orders, addressed above, are non-Abelian. Several Abelian orders have also been proposed at $\nu=5/2$ (for a review, see Refs. \onlinecite{foot0,x19-Nature-2018}). All of them exhibit an integer quantized thermal conductance. For example, the $K=8$ state has $K=3\kappa_0$, the $331$ state has $K=4\kappa_0$, the $113$ state has $K=2\kappa_0$. The observed half-integer quantization of the heat conductance supports a non-Abelian topological order \cite{x19-Nature-2018}.

We assume below that the $5/2$ state is spin-polarized. Since the anti-Pfaffian state is polarized, any attempt to reconcile the data with that state must make such an assumption. Note also that almost all existing evidence points towards  a spin-polarized $5/2$ liquid.
This picture is supported by numerics ~\cite{x1-Morf,num-spin-1,num-spin-2,num-spin-3} and NMR experiments \cite{Stern-NMR,Tiemann-NMR}. On the other hand, an optical experiment
 \cite{Stern-optical} was interpreted as supporting zero polarization. This interpretation is in conflict with a subsequent optical experiment \cite{Wurstbauer2013}. 
The measurements of the activation gap $\Delta_{5/2}$ suggest a possibility of a spin transition at the electron density $\rho\sim 5\times 10^{10}$ cm$^{-2}$~\cite{Pan-transition, Samkharadze2017}. The relevant charge densities are several times higher in the experiment \cite{x19-Nature-2018}. This is consistent with the $100\%$ spin polarization. Recent experiments in ZnO \cite{Falson-ZnO} suggest that spin-polarized composite fermions with the orbital quantum numbers of the second Landau level are responsible for half-integer quantum Hall plateaus. This supports the picture of a spin-polarized 
$5/2$ state in GaAs. It is also not obvious how to generalize the proposed non-Abelian states to an unpolarized system. Thus, the observed half-integer thermal conductance also supports a polarized quantum Hall liquid. Finally, spin polarization \cite{xnew-composite} of the composite fermion Fermi sea near $\nu=5/2$ is another evidence in favor of a polarized $5/2$ state.

\subsection{The upper spin branch of the first Landau level}

The states of the upper spin branch of the lowest Landau level ($1<\nu<2$) can be understood as particle-hole conjugates of QHE liquids from the lower spin branch ($\nu<1$). In this picture, one starts with two filled Landau levels and adds holes. Equivalently, one introduces composite fermions with the total density equal to that of the holes. 

The two filled Landau levels have zero spin polarization. The polarization of a fractional quantum Hall liquid depends on the spin of composite fermions. The latter is determined by the competition of the Zeeman and cyclotron  energies. A strong magnetic field, parallel to the 2D electron gas, increases the Zeeman energy and favors a spin-polarized state. 

\begin{figure}[htb]
\includegraphics[width=3.0 in]{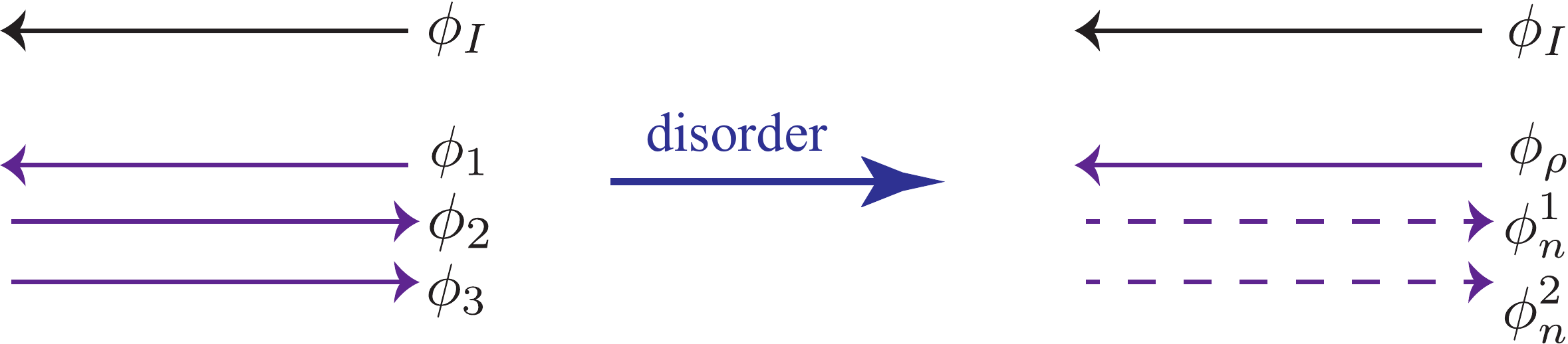}
\caption{(color online) Edge structure for the $\nu=8/5$ state. 
One integer edge channel $\phi_I$ separates $\nu=0$ from $\nu=1$.
The fractional edge separates $\nu=1$ and $\nu=8/5$ and contains one downstream charged mode with the conductance $e^2/h$ and two upstream modes with the total conductance $2e^2/5h$. Disorder reorganizes the fractional edge into a single downstream charged mode $\phi_\rho$ of the conductance $3e^2/5h$ and two upstream neutral modes 
$\phi_n^1$ and $\phi_n^2$.}
\label{fig:B}
\end{figure}

The physics is similar for all states with $3/2<\nu<2$. We will focus below on $\nu=8/5$. We will assume that the quantum Hall liquid is spin-polarized. 
Such polarized $8/5$ state was
observed in Ref. 
\onlinecite{8-5-polarized}. 
In the $8/5$ state, the filling factor is $2/5$ for holes. This translates into the filling factor of $2$ for composite fermions. The edge structure includes one integer downstream edge mode, separating $\nu=0$ from $\nu=1$. Fractional edge modes separate $\nu=1$ from $\nu=8/5$. Their structure is the same as in the $3/5$ state. In the absence of disorder, there is a downstream fractional mode of the conductance $e^2/h$ and two upstream modes of the combined conductance $2e^2/5h$. Disorder modifies this picture, as discussed in 
Section~\ref{sec:8/5_prediction} and illustrated in Fig. \ref{fig:B}. This, however, has no effect on the equilibrium thermal conductance, which is determined by the difference 
of the numbers of the upstream and 
downstream modes: $K\sim (2-2)=0$. We predict that the experiments of the type \cite{x19-Nature-2018,x24-Nature-2017} should show a very different $K_{8/5}=2\kappa_0$.

\begin{table*} [htb] 
\begin{center}
\begin{tabular}{| l | | l | c | l |}
\hline
\quad\quad\quad FQH state & ~integer downstream modes~ & ~fractional downstream modes~ & ~fractional upstream modes~ 
\\ \hline\hline
~$\nu=7/3$~ & ~~~~~~1C + 1N ; $K=2\kappa_0$~  & ~1C   ; $K=\kappa_0$~ & ~~~~~~~none ; $K=0$
\\ \hline
~$\nu=8/3$~ & ~~~~~~1C + 1N ; $K=2\kappa_0$~  & ~1C   ; $K=\kappa_0$~ & ~~~~~~~~~1N ; $K=\kappa_0$
\\ \hline
~$\nu=8/5$~ & ~~~~~~1C + 0N ; $K=\kappa_0$  & ~1C   ; $K=\kappa_0$~ & ~~~~~~~~~2N ; $K=2\kappa_0$
\\ \hline 
~$\nu=5/2$ (Pfaffian)~ & ~~~~~~1C + 1N ; $K=2\kappa_0$~ & ~1C+1M ; $K=1.5\kappa_0$~  & ~~~~~~~none  ; $K=0$
\\ \hline
~$\nu=5/2$ (anti-Pfaffian)~ & ~~~~~~1C + 1N ; $K=2\kappa_0$~ & ~1C ; $K=\kappa_0$~  & ~~~~~~~~~3M  ; $K=1.5\kappa_0$~
\\ \hline
~$\nu=5/2$ (PH-Pfaffian)~ & ~~~~~~1C + 1N ; $K=2\kappa_0$~ & ~1C ; $K=\kappa_0$~ & ~~~~~~~~~1M  ; $K=0.5\kappa_0$~
\\ \hline
\end{tabular}
\caption{Edge structures. The numbers of the Bose charged (C), Bose neutral (N), and Majorana neutral (M) modes with their combined thermal conductances.}
\label{tab:1}
\end{center}
\end{table*}

\section{Edge theory at $\nu=8/3$}
\label{sec:edge_8/3}

The purpose of this section is a detailed analysis of the edge structure at $\nu=8/3$. The lessons also apply at $\nu=8/5$ and $5/2$. 
We start with a review of experimental data at $\nu=8/3$. The data does not show clear quantization of $K$.  Our main goal is to explain that observation.

\subsection{Thermal conductance data}

If all $N$ chiral edge channels run in the same downstream direction, the thermal conductance \cite{x22-KF-eq} of an integer or Abelian fractional QHE state equals $N\kappa_0 T$. If $N_u$ upstream and $N_d$ downstream channels coexist, the absolute value of the thermal conductance is expected \cite{x22-KF-eq} to equal the difference $|N_u-N_d|\kappa_0 T$ of the contributions from the up- and down-stream modes. This was observed \cite{x24-Nature-2017} at $\nu=3/5$, where $N_d=1$, $N_u=2$, and $KT=(1.04\pm 0.03)\kappa_0 T$, and at $\nu=4/7$, where $N_d=1$, $N_u=3$, and $KT=(2.04\pm 0.05)\kappa_0T$. At the same time, the $\nu=2/3$ plateau shows \cite{x24-Nature-2017} $KT=0.33\kappa_0 T$ at the bath 
temperature $T_0=10~$mK, while $N_d=N_u=1$. A much greater deviation of $K$ from $(N_u-N_d)\kappa_0$ at $\nu=2/3$ than at $\nu=3/5$ and $4/7$ reflects a different temperature profile along the edge \cite{x24-Nature-2017}. When $N_u>N_d$, the temperature $T_u$ of the upstream modes remains constant along most of the edge in the absence of heat loses to the bulk \cite{x24-Nature-2017,so-2018}. The temperature of the minority downstream modes is also $T_u$ beyond an equilibration-length distance from their source, and the finite-size correction to $K$ rapidly vanishes as a function of the edge length $L$ in large systems. On the other hand, at $N_u=N_d$, the temperatures of the upstream and downstream modes are approximately the same in each point and change continuously along the edge \cite{x24-Nature-2017}. The finite size correction to $K$ exhibits a slow algebraic dependence on $L$. The $8/3$ edge contains 2 downstream integer modes in addition to one up- and one down-stream fractional mode. The expected $K=2\kappa_0$. Yet, the observed \cite{x19-Nature-2018} $K=(2.19\pm 0.03)\kappa_0$ at $T_0=11~-~12~$mK deviates much more from the equilibrium value than at $\nu=3/5$ and $4/7$. 

As discussed above, the $8/3$ state is believed to differ from the $2/3$ state only by two filled Landau levels \cite{8-3-2011}. Hence, if one assumed that the integer and fractional modes did not interact, the observed large deviation of $K$ from $2\kappa_0$ could be explained by the same physics as at $\nu=2/3$. Such explanation fails because of the long-range Coulomb force between the integer and fractional channels. Naively, strong Coulomb interaction results in the same temperature for all modes as at $\nu=3/5$ and $4/7$. Yet, this is not the case at $\nu=8/3$: Appropriately chosen collective modes interact weakly and do not equilibrate. In the next subsection, we find such collective modes.

\subsection{Edge action}

The filled lowest Landau level gives rise to two integer charged modes with opposite spins. The modes interact via Coulomb repulsion. Besides, there is random interaction between the modes. The Lagrangian density of the two integer channels
\begin{eqnarray}
L_{\rm int}=\sum_{i=\uparrow,\downarrow}
\frac{\partial_x\phi_i\left(\partial_t-v_i\partial_x\right)\phi_i}{4\pi}
-\frac{w_{\uparrow\downarrow}}{4\pi}\partial_x\phi_\uparrow\partial_x\phi_\downarrow
+L_{\rm r}^{\uparrow\downarrow}
\end{eqnarray}
Here, $\phi_\uparrow$ and $\phi_\downarrow$ denote the two Bose charged modes with the opposite spin projections and the charge densities $e\partial_x\phi_{\uparrow,\downarrow}/2\pi$.
A factor of $\hbar$ is  absorbed into $L_{\rm int}$.
$w_{\uparrow\downarrow}$ describes the strength of the Coulomb interaction. Disorder introduces a random density-density interaction 
$L_{\rm r}^{\uparrow\downarrow}$. 
We neglect spin-flip electron tunneling between the modes. This is justified by the weakness of the spin-orbit coupling. See Ref. \onlinecite{eq1-spin} for a discussion of the large 
spin equilibration length.

\begin{figure}[htb]
\includegraphics[width=3.0 in]{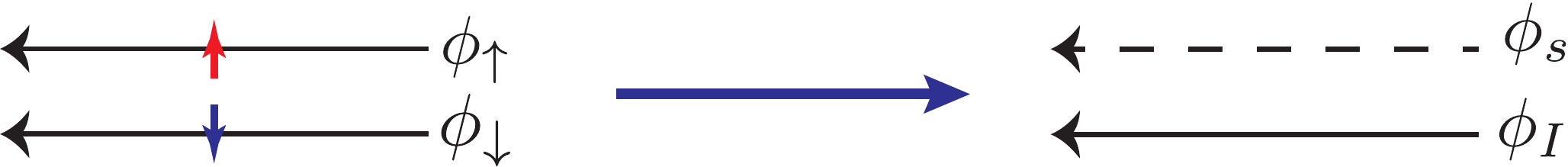}
\caption{(color online) Spin-charge separation in the integer channels at $2<\nu<3$. The charged mode 
$\phi_I$ has the conductance $2e^2/h$, whereas the spin mode $\phi_s$ is neutral.}
\label{fig:C}
\end{figure}

It is convenient to introduce a total integer charged mode $\phi_I=\phi_\uparrow+\phi_\downarrow$ and a spin mode $\phi_s=\phi_\uparrow-\phi_\downarrow$.
The Lagrangian density of the integer edge becomes
\begin{align}
\label{dima-new-2}
L_{\rm int}=
&\frac{1}{8\pi}\partial_x\phi_I\left(\partial_t-v_I\partial_x\right)\phi_I
+\frac{1}{8\pi}\partial_x\phi_s\left(\partial_t-v_s\partial_x\right)\phi_s
\nonumber \\
&-\frac{w_{sI}}{4\pi}\partial_x\phi_s\partial_x\phi_I
+L_{\rm r}^{s},
\end{align}
where the random interaction

\begin{eqnarray}
\label{dima-new-3}
L^s_{\rm r}
=\frac{1}{4\pi}\eta(x) \partial_x\phi_I\partial_x\phi_s
\end{eqnarray} 
with $\eta(x)$ being a random function of the coordinate. 
Random contributions to the self-interactions of the modes are omitted since they do not change the physics qualitatively.
The integer edge channels are illustrated in Fig.~\ref{fig:C}.

\begin{figure}[htb]
\includegraphics[width=3.0 in]{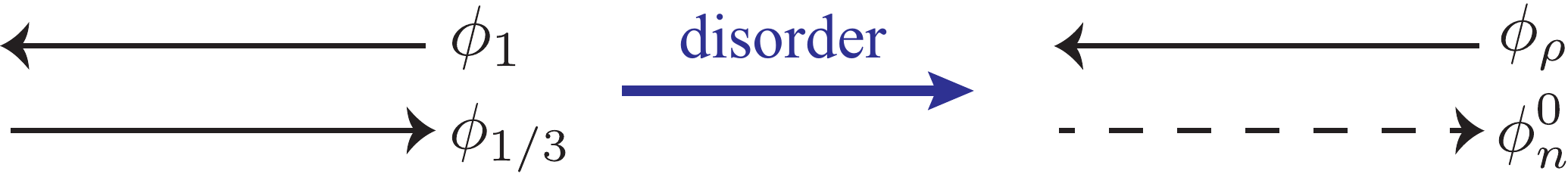}
\caption{Disorder effect on the fractional edge at $\nu=8/3$. The fractional charged mode $\phi_\rho$ has the conductance $2e^2/3h$, whereas the upstream mode 
$\phi_n^0$ is neutral.}
\label{fig:D}
\end{figure}

The fractional edge channels are the same as at $\nu=2/3$. There is one downstream charged mode $\phi_1$ with the conductance $e^2/h$ and an upstream charged mode $\phi_{1/3}$ with the conductance $e^2/3h$. In the absence of interaction, the total electrical conductance is $4e^2/3h$ \cite{KFP}. Kane, Fisher, and Polchinski 
observed that the Coulomb interaction between these two modes 
cannot ensure the correct quantized Hall conductance of $2e^2/3h$~\cite{KFP} and that random tunneling due to impurities is responsible for the observed quantization. 
Thus, the Lagrangian density takes the form~\cite{wen-book}:
\begin{align}
\label{dima-new-4}
L_{\rm f}=
&\frac{1}{4\pi}
\partial_x\phi_1
\left(\partial_t-v_1\partial_x\right)\phi_1
-\frac{3}{4\pi}\partial_x\phi_{1/3}
\left(\partial_t+v_{1/3}\partial_x\right)\phi_{1/3}
\nonumber \\
&-\frac{w_{\rm f}}{4\pi}\partial_x\phi_1\partial_x\phi_{1/3}
+L_{\rm r}^{\rm f},
\end{align}
with $\hbar$ being absorbed into $L_{\rm f}$ and with $L_{\rm r}^{\rm f}$ describing disorder.
The random tunneling term $L_{\rm r}^{\rm f}\sim \exp(i\phi_1+3i\phi_{1/3})$ is relevant in the renormalization group sense \cite{KFP} 
in the presence of strong Coulomb interaction and drives the edge theory  towards a fixed point with one upstream 
neutral mode $\phi_n^0$ and one downstream charged mode $\phi_\rho$ of the conductance $2e^2/3h$ (Fig.~\ref{fig:D}). The corresponding Lagrangian 
density \cite{KFP}
\begin{align}
\label{dima-new-5}
L_{\rm f}=
&\frac{3\partial_x\phi_\rho
\left(\partial_t-v_\rho\partial_x\right)\phi_\rho}{8\pi}
-\frac{\partial_x\phi_n^0
\left(\partial_t+v_n^0\partial_x\right)\phi_n^0}{4\pi}
+L^n_{\rm r},
\end{align}
where the random interaction of the two modes \cite{KF-randomness}

\begin{eqnarray}
\label{dima-new-6}
L^n_{\rm r}
=\frac{\partial_x\phi_\rho}{4\pi}\left[\xi(x)\partial_x\phi_n^0 
+\{\zeta(x)\exp(i\sqrt{2}\phi_n^0)+{\rm H.c.} \}\right]
\end{eqnarray}
with a random real $\xi(x)$ and a random complex $\zeta(x)$. 
The residual non-random interaction between $\phi_\rho$ and $\phi_n^0$ is an irrelevant perturbation \cite{KF-randomness} and does not alter the physics of the fractional edge. 
We drop it in the following discussion.

A combination of the Lagrangians for the integer (\ref{dima-new-2}) and fractional (\ref{dima-new-5}) channels yields the total Lagrangian density on the 
$\nu=8/3$ edge:
\begin{align}
\label{1}
L_{8/3}
=
&\frac{3\partial_x\phi_\rho\left(\partial_t-v_\rho\partial_x\right)\phi_\rho}{8\pi}
+\frac{\partial_x\phi_I\left(\partial_t-v_I\partial_x\right)\phi_I}{8\pi}
\nonumber \\
&+\frac{\partial_x\phi_s\left(\partial_t-v_s\partial_x\right)\phi_s}{8\pi}
-\frac{\partial_x\phi_n^0\left(\partial_t+v^0_n\partial_x\right)\phi_n^0 }{4\pi}
\nonumber \\
&-\frac{w_{\rho I}}{4\pi}\partial_x\phi_\rho\partial_x\phi_I
-\frac{w_{sI}}{4\pi}\partial_x\phi_s\partial_x\phi_I
\nonumber \\
&+L_{\rm r}^n+L_{\rm r}^s,
\end{align}
where $e\partial_x\phi_\rho/2\pi$ and $e\partial_x\phi_I/2\pi$ are the charge densities on the fractional and integer edges respectively, $\frac{1}{2}\frac{\partial_x\phi_s}{2\pi}$ is the spin density on the integer edge, $\phi_n^0$ is the upstream neutral mode, $w_{\rho I}$ and $w_{sI}$ describe nonrandom intermode interactions, $L^{n,s}_{\rm r}$ describe disorder and are given by Eqs. (\ref{dima-new-3}) and (\ref{dima-new-6}), and $\hbar$ is absorbed into $L_{8/3}$. We neglect the interaction of the modes $\phi_s$ and $\phi_n^0$ since 
they are spatially separated \cite{CCG,GH} and neutral, and hence do not participate in the long-range Coulomb interaction. 
The action (\ref{1}) also neglects tunneling between the integer and fractional channels. This is motivated by the large distance between the integer and fractional channels. For a discussion of large equilibration lengths between inner channels and outer integer channels, see Refs. \onlinecite{eq2-hirai,eq3-deviatov}.
The discussion below shows that other random and nonrandom interactions of the modes do not qualitatively change the results, and we do not include them in (\ref{1}).

The integer-mode interaction $w_{sI}$ depends on the asymmetry between the spin-up and -down integer channels and vanishes in a symmetric system. Since we deal with etched edges far away from the gates \cite{x19-Nature-2018}, the energy of the charged modes is dominated by the long-range Coulomb interaction. Hence, the coefficients $v_\rho$, $v_I$, and $w_{\rho I}$ are not independent. The three terms these coefficients define in the action combine approximately into $-\frac{1}{4\pi}\frac{3}{8}v_c(\partial_x\phi_c)^2$, where 
$e\partial_x\phi_c/2\pi=e\partial_x(\phi_\rho+\phi_I)/2\pi$ is the total charge density on the edge, and $v_c$ is the speed of the overall charged mode. We thus rewrite the action in terms of the charged mode $\phi_c$ and a new neutral mode $\phi_n=\sqrt{3}\phi_\rho-\phi_I/\sqrt{3}$:
\begin{align}
\label{2}
L_{8/3}
=&\frac{3\partial_x\phi_c\left(\partial_t-v_c\partial_x\right)\phi_c}{32\pi}
+\frac{3\partial_x\phi_n\left(\partial_t-v_n\partial_x\right)\phi_n}{32\pi}
\nonumber \\
&+\frac{\partial_x\phi_s\left(\partial_t-v_s\partial_x\right)\phi_s}{8\pi}
-\frac{\partial_x\phi_n^0\left(\partial_t+v^0_n\partial_x\right)\phi_n^0}{4\pi}
\nonumber \\
&-\frac{w_{n c}}{4\pi}\partial_x\phi_n\partial_x\phi_c
-\frac{w_{sI}}{16\pi}\partial_x\phi_s\partial_x(3\phi_c-\sqrt{3}\phi_n)
\nonumber \\
&+L_{\rm r}^n+L_{\rm r}^s,
\end{align}
where
\begin{equation}
\label{3}
L^n_{\rm r}\sim\partial_x\phi_\rho=\partial_x(A_n\phi_n+A_c\phi_c)
\end{equation}
with $A_n=\sqrt{3}/4$, $A_c=1/4$. The interaction $w_{n c}\ll v_c$ can be eliminated by one more change of variables: a small rotation in the two-dimensional space spanned by the variables $\phi_{n,c}$. We ignore $w_{n c}$ below.

\subsection{Fast charged mode}

To estimate the speed of the charged mode, we observe that the main contribution to $v_c$ comes from the long-range electrostatic repulsion. Consider a uniform charge density $\rho=e\partial_x\phi_c/2\pi$ on an edge segment of length $l\sim d$, where $d$ is the distance to the screening gates. The energy
\begin{align}
\label{4}
E=\frac{\hbar v_c l}{4\pi\nu}(\partial_x\phi_c)^2
&\approx\frac{\rho^2}{\epsilon^*}\int_w^l dx \int_0^{x-w} \frac{dy}{x-y}
\nonumber \\
&\sim(\rho^2 l/\epsilon^*)\ln\frac{d}{w},
\end{align}
where $w$ is the edge width, $\nu$ is the filling factor, and the effective dielectric constant $\epsilon^*=(\epsilon+1)/2\approx 7$ since 
$d\gg({\rm the~depth~of~the~electron~gas~under~the~surface})$. The typical distance $d$ from the gates is on the order of tens of microns. The estimate in equation~\eqref{4} requires $L_T=\hbar v_c/k_B T\gtrsim d$. In our case $L_T\sim L$. Estimates in the spirit of Refs. \onlinecite{CCG,GH} suggest that the edge is not much wider than a micron. Experiment supports \cite{edge-width-1,edge-width-2,baer} $w\sim ~{\rm hundreds~nm}$ at $\nu=5/2$ and likely  also \cite{baer} at $\nu=8/3$. Hence, $v_c\sim 5\times 10^7~{\rm cm/s}$. Even higher velocities $\approx 2\times 10^8~{\rm cm/s}$
were reported \cite{kumada-v} at $\nu\sim 2$. The neutral-mode velocities $v_{n,s}$ are much lower. The spin-mode velocity $v_s=4.6\times 10^6~{\rm cm/s}$ was reported in 
Ref. \onlinecite{feve}. A similar value was predicted by numerics \cite{kun-yang-2-3} for the neutral-mode velocity at $\nu=2/3$. Thus, we expect that the charged mode is an order of magnitude faster than the neutral modes \cite{hashisaka}.

\section{Equilibration length} 
\label{sec:length}

The rate of the energy exchange between two modes is set by the equilibration length $l_e$. The heat flow between the segments of length $l_e$ 
of the two channels is $\kappa_0 T\Delta T$, where $\Delta T$ is their temperature difference. 
In this section we estimate $l_e$ for different pairs of channels at $\nu=8/3$. The analysis of other filling factors is essentially the same.

We only consider temperature equilibration between edge channels. Following the calculation of the Joule heat in the central Ohmic reservoir
\cite{x24-Nature-2017}, short edge segments \cite{charge-eq} on which voltage equilibrates are treated as parts of Ohmic contacts (Appendix \ref{app:voltage}).

We first estimate $l_e$ with scaling analysis and then reproduce the answer with a perturbative calculation.

\subsection{Scaling analysis}

The equilibration length $l_e^n$ between $\phi_n$ and $\phi_n^0$ in Eq. (\ref{2}) can be deduced from scaling. The energy exchange between the modes is due to disorder (\ref{3}). 
As discussed in Ref. \onlinecite{comment}, one expects that the disorder correlation length is not much greater than the thermal length $l_T\sim\hbar v_n^0/T$. 
We can thus consider delta-correlated disorder. For example, $\langle\zeta(x)\zeta(y)\rangle =W\delta(x-y)$. 
The non-random quadratic terms in Eq. (\ref{2}) are invariant under the action of renormalization group (RG). Disorder is irrelevant and satisfies the RG equation
\begin{eqnarray}
\frac{dW}{dl}=(3-2\Delta)W,
\end{eqnarray}
where $\Delta=2$ is the scaling dimension. The RG flow stops at the thermal length $l_T$. At that scale, $W_{\rm eff}\sim W l_T^{3-2\Delta}=W l_T^{-1}$. The energy current 
$\mathcal{J}_n$ between the channels
$\phi_n$ and $\phi_n^0$ on the edge segment of length $l_T$ is proportional (i) to $W_{\rm eff}$, (ii) to the square of the coupling constant $A_{n}$ in Eq. (\ref{3}), and (iii) to the tunneling density of states $D_{n}$ available for heat transfer into the $\phi_n$ channel. To find the density of states we fermionize \cite{bosonization} $\phi_n$. $\phi_n$ enters the disorder operator (\ref{3}) as $\partial_x\phi_n$. After fermionization, $\partial_x\phi_n$ reduces to a product of two fermion operators. Their speed $v_n$ is the same as for the original Bose-mode $\phi_n$. Hence, the relevant density of states $D_n\sim 1/v_n^2$.
Thus, the heat $\mathcal{J}_{n}\sim T\Delta T A_{n}^2 W_{\rm eff}/v_n^2$. The dependence of $\mathcal{J}_{n}$ on $v_n^0$ is unimportant below but can be easily found with fermionizing $\phi_n^0$:
$\mathcal{J}_{n}\sim 1/[v_n^0]^2$.
The equilibration length is defined by $\kappa_0T\Delta T\sim (l^{n}_e/l_T) \mathcal{J}_{n}$. Therefore
\begin{eqnarray}
\label{dima-new-7}
l^{n}_e\sim
\frac{\left( v_n^0 v_{n}\right)^2}{W A_{n}^2} T^{-2}.
\end{eqnarray}
Exactly the same calculation applies to the equilibration length $l_e^c$ between the charged mode $\phi_c$ and $\phi_n^0$. The only change in Eq. (\ref{dima-new-7}) consists in the substitution of the index $n$ with $c$ in all terms except $v_n^0$.

\subsection{Perturbation theory}

The equilibration length $l_e^n$ for the modes $\phi_n$ and $\phi_n^0$ can be estimated by computing the scattering rate for edge excitations in the second order perturbation 
theory in $L_{\rm r}^n$. Below, we focus on the second contribution in $L_{\rm r}^n$ (Eqs. \ref{dima-new-6}, \ref{3}): $A_n\partial_x\phi_n\left[\zeta(x)\exp(i\sqrt{2}\phi_n^0)+{\rm H.c.}\right]/(4\pi)$. The analysis of the other contribution is similar. We start with fermionizing the action~\cite{bosonization}. We first set $\sqrt{2}\phi_n^0=\theta_1-\theta_2$ and add to the action an auxiliary field $\tilde{\phi}_n^0=(\theta_1+\theta_2)/\sqrt{2}$ with the same velocity as that of $\phi_n^0$. We then introduce fermionic operators 
$\psi_1^{\dagger}=e^{i\theta_1}/\sqrt{2\pi\alpha}$ and 
$\psi_2=e^{-i\theta_2}/\sqrt{2\pi\alpha}$, where $\alpha$ is an ultraviolet cutoff. The Hamiltonian of the random interaction becomes
\begin{eqnarray}
H_{\rm r}=
-\frac{\hbar}{2} A_n 
\int\tilde{\zeta}(x)
\partial_x \phi_n
\psi_1^{\dagger}\psi_2 dx
+{\rm H.c.},
\end{eqnarray}
where $\tilde{\zeta}(x)=\alpha\zeta(x)$. We expand the field operators in the plane-wave basis:
\begin{align}
\phi_n(x)
&=\sqrt{\frac{8}{3}}
\sum_{q<0 }
\frac{1}{\sqrt{n}}
\left(e^{iqx}a_q + e^{-iqx}a_q^{\dagger}\right),
\\
\psi^{\dagger}_1(x)
&=\frac{1}{\sqrt{L}}\sum_{k_1}e^{-ik_1x}b_{k_1}^{\dagger},
\\
\psi_2(x)
&=\frac{1}{\sqrt{L}}\sum_{k_2}e^{ik_2x}c_{k_2},
\end{align}
where $q=-2n\pi/L$, and $L$ is the edge length. Then 
\begin{align}
\nonumber
H_{\rm r}
=\frac{i\pi\hbar}{L^2}\sqrt{\frac{8}{3}}A_n 
\sum_{\substack{n>0 \\ k_1, k_2}}
\sqrt{n}
&\left(
\tilde{\zeta}_{q-k_1+k_2} 
a_q b_{k_1}^{\dagger} c_{k_2} \right.
\\
&-\left. \tilde{\zeta}_{-q-k_1+k_2} 
a_q^{\dagger} b_{k_1}^{\dagger} c_{k_2}
\right)+{\rm H.c.}
\end{align} 
Following the discussion in Section \ref{sec:thermal_8/3}, we neglect $w_{sI}$ in the action (\ref{2}).

The Fermi golden rule yields the scattering rate from an initial state with the excitations of the momenta $q$ and $k_2$ to the final state with the excitation of the momentum $k_1$ as
\begin{align}
&R_{(q, k_2)\rightarrow k_1}
\nonumber \\
=&\frac{2\pi}{\hbar} 
\left|\frac{i\pi\hbar}{L^2}\sqrt{\frac{8}{3}}A_n \sqrt{n}\right|^2
\left(\overline{\zeta^2}L\right)
\left| \langle f |
a_q b_{k_1}^{\dagger} c_{k_2}
| i\rangle\right|^2
\nonumber \\
&\times 
\delta\left[\hbar v_n^0 (k_1- k_2)-\hbar v_n |q| \right]
\end{align}
In the above calculation, we substitute the correlation 
$\langle \tilde{\zeta}(x)\tilde{\zeta}(y)\rangle =\overline{\zeta^2}\delta(x-y)$
and observe that
\begin{eqnarray}
| \tilde{\zeta}_q |^2
=\int_0^L \int_0^L e^{iqx} e^{-iqy} \tilde{\zeta}(x)\tilde{\zeta}(y) dx dy
=\overline{\zeta^2}L
\end{eqnarray}
From the net scattering rate 
$R_{\rm net}=R_{(q, k_2)\rightarrow k_1}
+R_{(q, k_1)\rightarrow k_2}
-R_{k_1\rightarrow (q, k_2)}
-R_{k_2\rightarrow (q, k_1)}$,
the energy flux is calculated as
\begin{align}  \label{eq:flux}
\dot{\varepsilon}
=&~2\times
\frac{2\pi^3 \hbar^2\overline{\zeta^2}A_n^2}{\hbar L^3}
\left(\frac{8}{3}\right)
\left(\frac{L}{2\pi}\right)^4
\left(\frac{1}{\hbar v_n}\right)
\left(\frac{1}{\hbar v_n^0}\right)^2
\nonumber \\
&
\int_{0}^{\infty} 
\left(\epsilon_q |q|\right)  d\epsilon_q
\int_{-\infty}^{\infty} 
d\epsilon_{k_1} 
dk_2
~\delta\left[( k_1- k_2)-\frac{v_n}{v_n^0}|q|\right]
\nonumber \\
&\times
\left[
\left|\langle f |
a_q b_{k_1}^{\dagger} c_{k_2}
| i\rangle\right|^2
-\left|\langle f |
a_q^{\dagger} b_{k_1} c_{k_2}^{\dagger}
| i\rangle\right|^2\right],
\end{align}
where $\epsilon_q=\hbar v_n |q|$, $\epsilon_{k_i}=\hbar v_n^0 k_i$ are the energies of the modes $\phi_n$ and $\psi_i$, respectively. 
The final result is
\begin{align} \label{eq:flux_result}
\dot{\varepsilon}
&= \frac{2\pi^3 L}{45\hbar^5}\left(\hbar^2\overline{\zeta^2}\right)
\left(\frac{A_n}{v_n v_n^0}\right)^2
\left[\left(k_B T_n\right)^4
-\left(k_B T_n^0\right)^4\right]
\nonumber \\
&\sim\frac{1}{l^n_e}T\Delta T,
\end{align}
where $T_n$ and $T_n^0$ are the temperatures of the two modes, 
and the equilibration length $l^n_e$ sets a length scale for the modes to equilibrate:
\begin{eqnarray}
l_e^n 
\sim\left(\frac{v_n v_n^0}{A_n}\right)^2
\frac{1}{\overline{\zeta^2}} T^{-2}
\end{eqnarray}
Since $l_e^n$ is proportional to the speed squared, a faster edge mode requires a longer 
$l_e^n$ to achieve thermal equilibration with $\phi_n^0$.

A similar calculation yields the equilibration length $l_e^c$ for $\phi_c$ and $\phi_n^0$. At a small $w_{sI}$, one finds
\begin{equation}
\label{5}
\frac{l_e^c}{l_e^n}=\left(\frac{A_n}{A_c}\frac{v_c}{v_n}\right)^2.
\end{equation}
According to the phenomenological theory \cite{x24-Nature-2017}, $l_e^n$ is several times shorter than the total edge length $L$ in the experiments \cite{x19-Nature-2018,x24-Nature-2017}.
Thus, $l_e^c\gg L$. The above estimate should be taken with care because of the limitations of the phenomenological theory
(Appendix \ref{app:losses})
and because the 
perturbative calculation of $l_e^n$ is only valid, if $l_e^n\gg \hbar~ {\rm min}(v_n, v_n^0)/k_BT$. Nevertheless, a very large ratio (\ref{5}) suggests that 
the energy exchange between $\phi_n^0$ and $\phi_c$ is negligible. In the absence of the spin mode, this would justify the same physical picture of the thermal transport by the 
channels $\phi_n$ and $\phi_n^0$ as in the theory of the $2/3$-edge. Before focusing on the spin mode, we will consider the upper spin branch of the first Landau level, where an integer spin mode does not exist.

\section{$\nu=8/5$}
\label{sec:8/5_prediction}

We concentrate on a spin-polarized state at $\nu=8/5$~\cite{8-5-polarized}. Similar analysis applies at the other filling factors $\nu=2-n/(2n+1)$.
The edge includes an integer mode and fractional modes of the same nature as in the lower spin branch at $\nu=\frac{n+1}{2n+1}$. The latter filling factors were investigated 
by Kane and Fisher \cite{KF-randomness} who found that disorder reorganizes the edge into a single downstream charged mode and $n$ upstream neutral modes of the same velocity.
The edge structure at $\nu=8/5$ is shown in Fig.~\ref{fig:B}. The Lagrangian density $L_{8/5}$ differs from (\ref{1}) by the absence of $\phi_s$
and the presence of two upstream neutral modes 
$\phi^1_n$ and $\phi^2_n$:
\begin{align}
\label{6}
L_{8/5}
=&\frac{5\partial_x\phi_\rho\left(\partial_t-v_\rho\partial_x\right)\phi_\rho}{12\pi}+\frac{\partial_x\phi_I\left(\partial_t-v_I\partial_x\right)\phi_I}{4\pi} 
\nonumber \\
&-\sum_{k=1,2}
\frac{\partial_x\phi_n^k\left(\partial_t+v^0_n\partial_x\right)\phi_n^k}{4\pi} 
-\frac{w_{\rho I}}{4\pi}\partial_x\phi_\rho\partial_x\phi_I
\nonumber \\
&+L_{\rm r}^n,
\end{align}
where $L_{\rm r}^n\sim\partial_x\phi_\rho$ describes the random interaction of up- and down-stream fractional modes \cite{KF-randomness}. The interaction of the two upstream modes is unimportant since it does not transfer energy between up- and down-stream channels. After introducing the overall charged mode $\phi_c=\phi_\rho+\phi_I$ and a downstream neutral mode $\phi_n=\sqrt{5/3}\phi_\rho-\sqrt{3/5}\phi_I$, one finds
\begin{align} \label{7}
L_{8/5}
=&\frac{5}{32\pi}
[\partial_x\phi_c\left(\partial_t-v_c\partial_x\right)\phi_c
+\partial_x\phi_n\left(\partial_t-v_n\partial_x\right)\phi_n] 
\nonumber \\
&-\sum_{k=1,2}
\frac{\partial_x\phi_n^k\left(\partial_t+v^0_n\partial_x\right)\phi_n^k}{4\pi} 
-\frac{w_{n c}}{4\pi}\partial_x\phi_c\partial_x\phi_n
\nonumber \\
&+L_{\rm r}^n,
\end{align}
where $w_{n c}$ can be ignored for the same reasons as in Eq. (\ref{2}).
The random interaction
\begin{eqnarray}
L_{\rm r}^n
\sim \left(\frac{3}{8}\partial_x\phi_c+\frac{\sqrt{15}}{8}\partial_x\phi_n\right)
\bm {V}\left[\phi_n^1,\phi_n^2\right],
\end{eqnarray}
where $\bm{V}\left[\phi_n^1,\phi_n^2\right]$ is a linear combination of five bosonic operators \cite{KF-randomness}
\begin{eqnarray}
\begin{aligned}
\bm{V}_1\left[\phi_n^1,\phi_n^2\right]
&=e^{i\sqrt{2}\phi_n^1}
\\
\bm{V}_2\left[\phi_n^1,\phi_n^2\right]
&=e^{i\frac{1}{\sqrt{2}}\phi_n^1+i\sqrt{\frac{3}{2}}\phi_n^2}
\\
\bm{V}_3\left[\phi_n^1,\phi_n^2\right]
&=e^{-i\frac{1}{\sqrt{2}}\phi_n^1+i\sqrt{\frac{3}{2}}\phi_n^2}
\\
\bm{V}_{4,5}
&=\partial_x \phi_n^{1,2}.
\end{aligned}
\end{eqnarray}
With two up- and two down-stream modes, one expects zero thermal conductance for a long edge in equilibrium.

After refermionizing $\phi_n^1$ and $\phi_n^2$, the equilibration length can be evaluated similarly to the $\nu=8/3$ state. We observe that Eqs. (\ref{5}) and (\ref{3}) apply at $\nu=8/5$ with $A_c/A_n=\sqrt{3/5}$. Thus, at a sufficiently large $d/w$, $l_e^c\gg l_e^n$. If the sample length satisfies $l_e^c\gg L\gg l_e^n$, then the charged mode $\phi_c$ decouples from the rest of the modes. The rest of the modes equilibrate. The thermal conductance becomes the sum of the contribution of a single mode $\phi_c$ and the contribution of the other three modes. Both contributions equal one quantum, and so the predicted
\begin{equation}
\label{8}
K_{8/5}=2\kappa_0.
\end{equation}
This result assumes ideal contacts: Each chiral mode in Eq. (\ref{7}) emanates from an Ohmic contact with the temperature of the contact. For non-ideal contacts, we expect $K_{8/5}$ between $2\kappa_0$ and 0.

\section{Back to $\nu=8/3$}
\label{sec:thermal_8/3}

To understand the $\nu=8/3$ physics, we need to consider the interaction of the spin mode $\phi_s$ with the other modes. We do not expect its interaction with the fast charged mode to play much role. Thus, we focus on the random and nonrandom interactions of $\phi_s$ and $\phi_n$, Eq. (\ref{2}). It is hard to estimate the interactions theoretically. We will try to extract constraints on the interaction from the experimental data. 

A simple phenomenological model from Appendix~\ref{app:pheno-model} shows that any strength of the random interaction $L_{\rm r}^s$ is compatible with the data at roughly the same interaction of the upstream fractional mode $\phi_n^0$ with $\phi_n$. It is unclear what happens at a strong nonrandom interaction $w_{sI}$. 
We argue below that the experimental data are consistent with a weak $w_{sI}$.

Let us focus on the nonrandom contributions to the Lagrangian (\ref{2}). The nonrandom terms that contain the fields $\phi_s$ and $\phi_n$ only are
\begin{align}
L^{(s,n)}_{8/3}
=&\frac{3\partial_x\phi_n\left(\partial_t-v_n\partial_x\right)\phi_n}{32\pi}
+\frac{\partial_x\phi_s\left(\partial_t-v_s\partial_x\right)\phi_s}{8\pi}
\nonumber \\
&+\frac{\sqrt{3}w_{sI}}{16\pi}\partial_x\phi_s\partial_x\phi_n
\end{align}
A weak $w_{sI}$ can be eliminated by the change of variables $\phi_n=\theta_n+\gamma \theta_s,$ $\phi_s=(\sqrt{3}/2)(\theta_s-\gamma\theta_n)$, where $\gamma=w_{sI}/[2(v_n-v_s)]$. This variable change generates from $L^n_{\rm r}$ (\ref{dima-new-6},\ref{2}) a random interaction of the new spin mode $\theta_s$ with the upstream mode $\phi_n^0$. From equation~\eqref{3}, one finds

\begin{align}
L^n_{\rm r}\sim
&{A_n} \partial_x\phi_n +A_c \partial_x\phi_c=
\nonumber \\
&
A_n\left[\partial_x\theta_n +\frac{w_{sI}}{2(v_n-v_s)}\partial_x\theta_s\right]+A_c\partial_x\phi_c.
\end{align}
The equilibration length $l_e^s$ can be computed in the same way as above. One finds
\begin{equation}
\label{9}
\frac{l_e^n}{l_e^s}=\left(\frac{w_{sI}}{2[v_n-v_s]}\frac{v_n}{v_s}\right)^2.
\end{equation}

A small $w_{sI}$ corresponds to little energy exchange between the spin mode and the rest of the system. This is compatible with the data since 
the observed imperfect quantization \cite{x19-Nature-2018} of $K_{8/3}$ can be explained by the decoupling of $\phi_s$ and $\phi_c$ from the rest of the modes.
A zero $w_{sI}$ implies symmetry between the integer edge modes $(\phi_I\pm\phi_s)$ with the opposite spin projections. We are not aware of a reason for such symmetry. 
At the same time, it is natural to expect approximate symmetry since the cyclotron gap considerably exceeds the Zeeman gap \cite{spin-dima-new}.
Moreover, the above interpretation of the experiment does not need a zero or very small $w_{sI}$. First, the small parameter is squared in Eq. (\ref{9}). Second, the effect of the energy exchange between the upstream and spin modes does not have to be negligible. It is possible that $(K_{8/3}-2\kappa_0)$ would be considerably greater than $0.19\kappa_0$ without such energy exchange. 
The value of $w_{sI}$ can be found experimentally by performing the experiments \cite{feve,hashisaka} at $\nu>2$.

\section{$\nu=5/2$} 
\label{sec:5/2}

What about $\nu=5/2$? The Lagrangian density for the anti-Pfaffian order is given by~\cite{x3-LRH,x4-LRNF}:
\begin{align}
\label{10}
L_{5/2}
=
&\frac{\partial_x\phi_\rho\left(\partial_t-v_\rho\partial_x\right)\phi_\rho}{2\pi}
+\frac{\partial_x\phi_I\left(\partial_t-v_I\partial_x\right)\phi_I}{8\pi} 
\nonumber \\
&+\frac{\partial_x\phi_s\left(\partial_t-v_s\partial_x\right)\phi_s}{8\pi}
+i\sum_{k=1}^3 \psi_k\left(\partial_t+v_\psi\partial_x\right)\psi_k
\nonumber \\
&-\frac{w_{\rho I}}{4\pi}\partial_x\phi_\rho\partial_x\phi_I
-\frac{w_{sI}}{4\pi}\partial_x\phi_s\partial_x\phi_I
\nonumber \\
&+L_{\rm r}^n+L_{\rm r}^s,
\end{align}
where $\psi_k$ are three Majorana fermions, and disorder $L_{\rm r}^n\sim i\sum_{k>l}\partial_x\phi_\rho(x)\psi_k(x)\psi_l(x)\zeta_{kl}(x)$ with a real  random $\zeta_{kl}$. 
$L_{\rm r}^n$ is quadratic in Majoranas since it must satisfy the Bose statistics.
The unimportant interaction of Majoranas is omitted. The same procedure as at $\nu=8/3$ and $8/5$ introduces new variables $\phi_c=\phi_\rho+\phi_I$ and $\phi_n=2\phi_\rho-\phi_I/2$.  Eqs. (\ref{3}) and (\ref{5}) hold now with $A_c/A_n=1/2$.
Again the large ratio $v_c/v_n$ leads to 
$l_e^c\gg l_e^n$, and hence the charged mode decouples from the rest of the channels.

Assume for a moment that the spin mode also decouples. Then the thermal conductance is the sum of two quanta from $\phi_s$ and $\phi_c$ plus the contribution of the remaining four modes. Three of them are upstream Majorana fermions with the thermal conductance of $0.5\kappa_0T$ per channel. If those four modes equilibrate with each other, their total thermal conductance $(3\times 0.5-1)\kappa_0T=0.5\kappa_0T$. This yields $K_{5/2}=2.5\kappa_0$ in agreement with the data at $T_0\sim 20~$mK. 

There are two problems with the above calculation. First, in contrast to the PH-Pfaffian hypothesis, it sheds no light on a much stronger temperature dependence
\cite{x19-Nature-2018,comment} of $K$ at $T_0\approx 10~$mK at $\nu=5/2$ than at all other filling factors. 
Second, there is no reason for the spin mode to completely decouple. Both problems can be 
solved by assuming that the interaction of the other modes with $\phi_s$ decreases the observed $K$ by $\Delta K_1\sim 0.2\kappa_0$ 
(see Appendix \ref{app:correction} for a discussion of this and other small corrections to the thermal conductance). 
Indeed, since the filling factor $5/2\approx 8/3$, we expect $w_{sI}$ to 
have similar values at $\nu=5/2$ and $8/3$ in the same sample. 
Thus, incomplete decoupling of the spin mode at $\nu=5/2$ is consistent with the picture of incomplete decoupling at $\nu=8/3$, addressed below Eq. (\ref{9}). 
In this picture, the observed $K_{5/2}=2.53\kappa_0$ implies partial equilibration of the Majorana modes with $\phi_n$ since 
such partial equilibration increases (Ref. \cite{x24-Nature-2017}, Appendix \ref{app:correction}) $K$ by some $\Delta K_2$.
Then the data can be explained by assuming that $\Delta K_2\approx \Delta K_1$. We are not aware of a reason for such cancellation. 
Yet, it is not impossible, given the limited amount and accuracy of the data. 
A strong temperature dependence of $K_{5/2}$ below 15 mK can then be explained by the breakdown of the cancellation between $\Delta K_1$ and $\Delta K_2$.

The breakdown reflects opposite temperature dependencies of $\Delta K_1$ and $\Delta K_2$ at low $T$. 
$\Delta K_1\sim L/l_e$ decreases at $T\rightarrow 0$ since $l_e$ diverges in that limit. 
At the same time, the growth of the equilibration length between $\phi_n$ and the Majorana modes at low temperatures implies the growth \cite{x24-Nature-2017} 
of the positive correction $\Delta K_2$. Thus, the combined correction $\Delta K_2-\Delta K_1$ increases at low $T$ in agreement with the data.

In addition to the fine-tuning of $\Delta K_{1,2}$, the above picture requires ideal contacts. This assumption can be tested by experimentally verifying the prediction (\ref{8}). At the same time, the PH-Pfaffian hypothesis explains the data without this assumption or any fine-tuning. Moreover, in contrast to the anti-Pfaffian picture, it sheds light on seemingly contradictory tunneling data \cite{x12-ZF,edge-width-2,baer,xilin1,xilin2} at $\nu=5/2$.

\section{Conclusions} \label{sec:conclusion}

Since $8/3\approx 5/2$, the observed imperfect quantization at $\nu=8/3$ may seem to put under question the results at $\nu=5/2$. Our mechanism solves this challenge. At the same time, 
only new experiments can fully answer the puzzle of the $5/2$ liquid. In particular, theory only allows  crude estimates of the coefficients in the edge action. Recent experimental work
\cite{feve,hashisaka} determined the coefficients in the edge action for integer QHE. It would be desirable to obtain similar information for fractional states too.

In conclusion, we predict that the QHE edges do not equilibrate in thermal conductance experiments with the setup \cite{x19-Nature-2018} in the upper spin branch of the first Landau level. In particular, the observed thermal conductance at $\nu=8/5$ is expected to be $2\kappa_0 T$, even though the equilibrium thermal conductance is 0. A similar mechanism explains the observed imperfect quantization of the thermal conductance at $\nu=8/3$. Under certain unlikely but not impossible assumptions, this mechanism might reconcile the observed thermal conductance at $\nu=5/2$ with the anti-Pfaffian hypothesis. The mechanism can be tested by increasing  the edge length since our assumptions do not apply to very long edges and $K$ is expected to decrease to the equilibrium value.

\begin{acknowledgments}

The authors  thank M. Banerjee, M. Heiblum, and V. Umansky for useful discussions. This research was supported in part by the National Science Foundation under Grant No. DMR-1607451.

\end{acknowledgments}

\appendix

\section{Voltage equilibration}
\label{app:voltage}

\begin{figure}[htb]
\includegraphics[width=3.4 in]{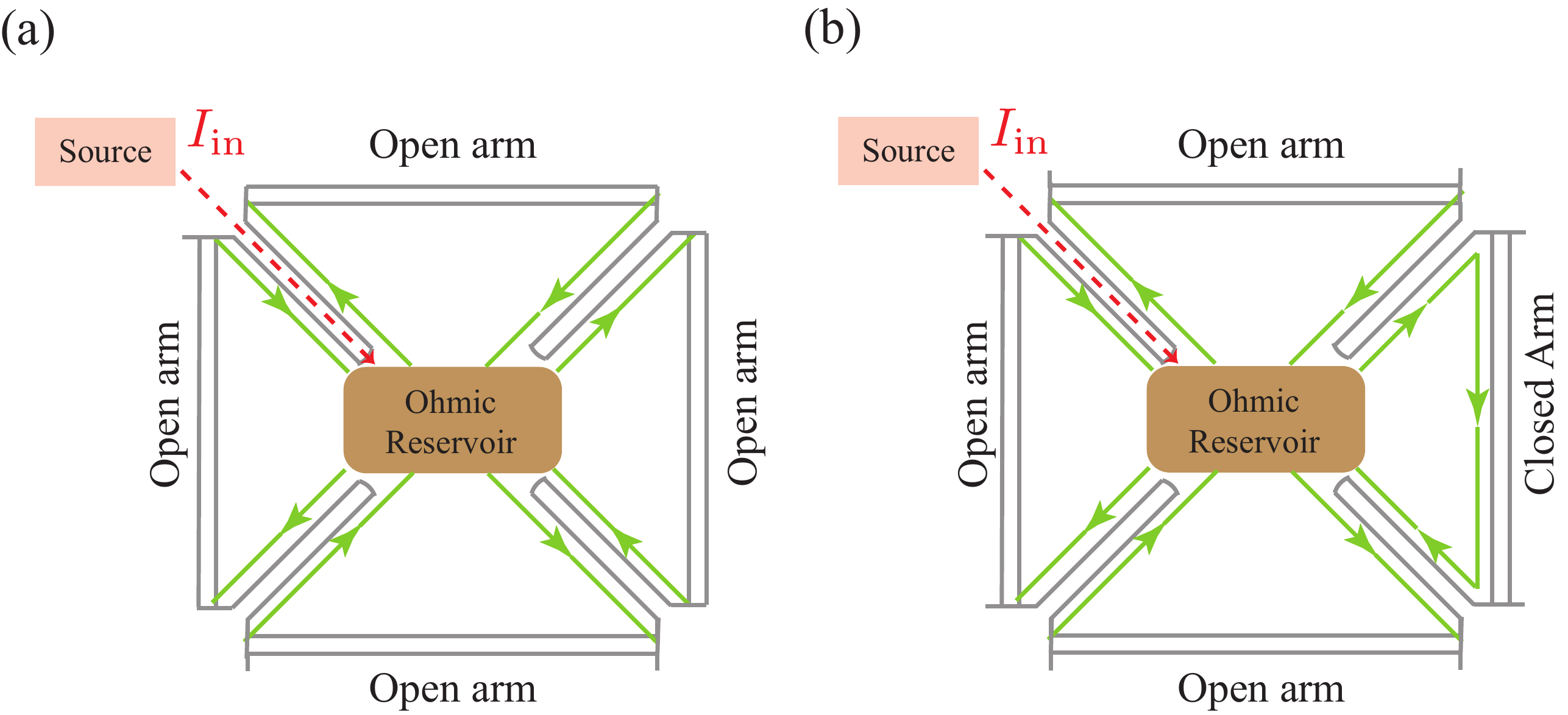}
\caption{(color online) A central Ohmic reservoir with $N=4$ arms. (a) All arms are open. (b) One of the arms is closed.}
\label{fig:app-dima-1}
\end{figure}

Refs. \onlinecite{x19-Nature-2018,x24-Nature-2017} compute the Joule heat in the central Ohmic reservoir in the following way. Let
$I_{\rm in}=GV$ be the current impinging on the central reservoir from the source (Fig. \ref{fig:app-dima-1}a). Here $V$ is the voltage bias and $G$ is the quantized Hall conductance of an edge. The impinging current carries the energy $E_{\rm in}\sim I_{\rm in}^2$. The current leaves the reservoir via $N$ arms. Each carries the current $I_{\rm out}=I_{\rm in}/N$.
This implies that the voltage of the central reservoir is $V/N$. The outgoing current carries the energy $E_{\rm out}\sim N I^2_{\rm out}=E_{\rm in}/N$. The energy difference $W=E_{\rm in}-E_{\rm out}$ is the Joule heat, dissipated in the reservoir.

\begin{figure}[htb]
\includegraphics[width=3.0 in]{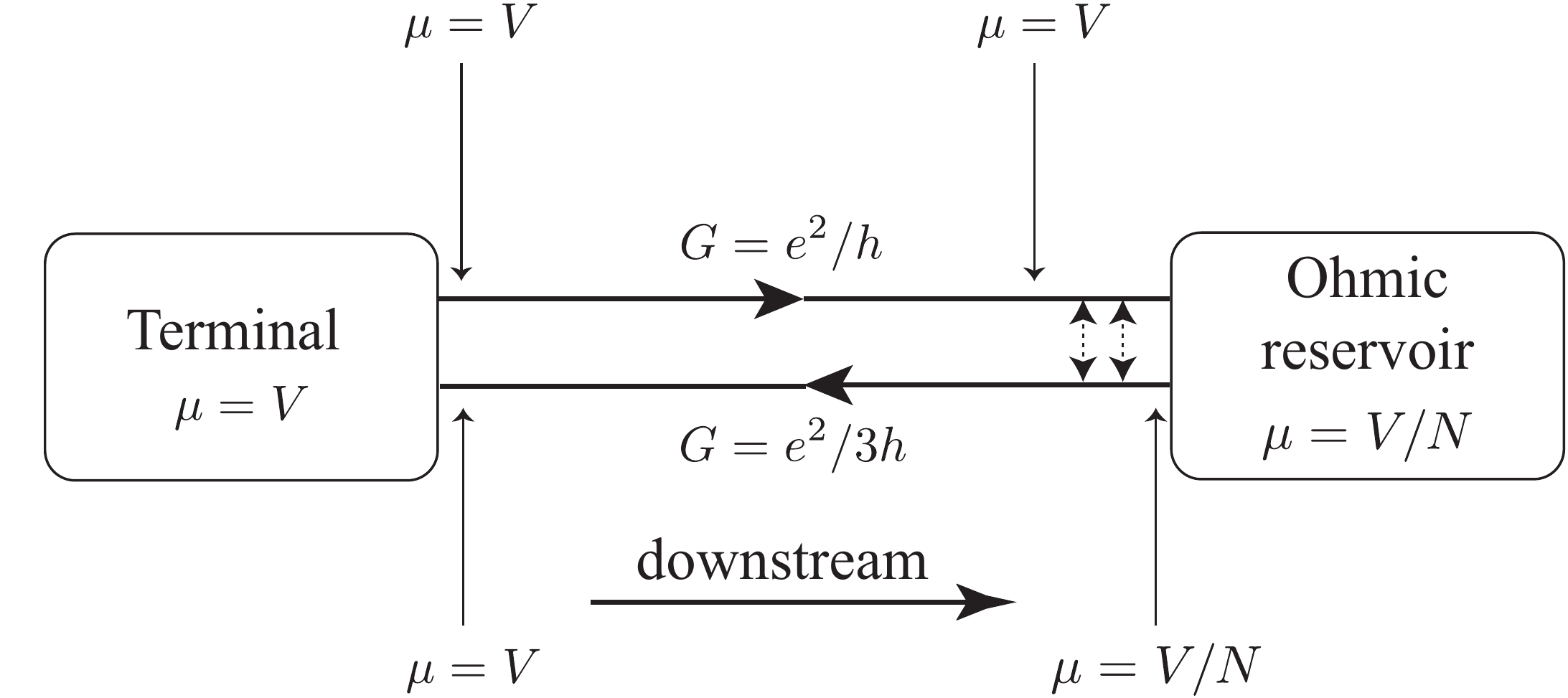}
\caption{Voltage equilibration between upstream and downstream charged modes. Dashed lines depict electron tunneling in the short region, where the chemical potential of the upstream and downstream channels equilibrate.}
\label{fig:app-dima-2}
\end{figure}

Such picture is clearly justified for chiral edges. We consider edges with contrapropagating channels. Up- and down-stream channels can have different chemical potentials near the central Ohmic reservoir as well as near the terminals  (Fig. \ref{fig:app-dima-2}). A common chemical potential is established through charge tunneling. The total Joule heat is the same as in the above calculation, but its portion is dissipated in the tunneling regions outside the Ohmic reservoirs, Fig. \ref{fig:app-dima-2}. This can be reconciled with Ref. \onlinecite{x19-Nature-2018}, if one assumes that all heat from the tunneling regions returns to the nearby Ohmic reservoirs. Such assumption is justified by the shortness of the voltage equilibration length which is much smaller than the reservoir size. The maximal equilibration length is observed for spin-unpolarized edges \cite{charge-eq} and is estimated as
$l_c\sim 2~\mu$m. We are interested in spin-polarized fractional edge channels in the second Landau level. A comparison with Ref. \onlinecite{charge-eq} suggests that for such edges $l_c$
is considerably lower than 2 $\mu$m and much shorter than the size of the central Ohmic reservoir.

\section{Bulk energy losses}
\label{app:losses}
Limitations of the phenomenological theory \cite{x24-Nature-2017} of heat transport include neglecting energy losses from the edge to the bulk. 

If all edge modes propagate downstream and all arms are open in 
the setup of Fig. \ref{fig:app-dima-1}, then it is justified to ignore the losses since bulk energy losses do not affect the observed heat conductance. 
Indeed, the experimental technique \cite{x24-Nature-2017} connects the energy outflow from the heated central Ohmic reservoir with the reservoir temperature. 
It does not matter for the interpretation of the data, if the heat, emitted by the central reservoir, reaches the terminals at the opposite ends of the edges 
or is lost to the bulk. It is only important that no thermal energy returns to the central Ohmic reservoir along the edges. This is, of course, guaranteed for chiral edges. 

The physics changes for non-chiral edges with upstream modes because upstream modes can return thermal energy back to the central Ohmic reservoir. A version of the phenomenological theory \cite{so-2018} shows that the bulk losses increase the observed thermal conductance of an open arm. Yet, there is no way to estimate the strength of this effect from the phenomenological theory. 

Fortunately, the bulk losses have an additional effect on the thermal conductance, when some of the arms are closed. This effect has not been discussed in the literature. 
In the absence of bulk losses, a closed arm  is irrelevant for thermal transport: 
All heat that enters the arm from the central reservoir returns to the reservoir (see Fig. \ref{fig:app-dima-1}b). At the same time, bulk losses from a closed arm decrease the returning thermal energy. 
This increases the thermal conductance of a device with closed arms. Curiously, this {\it decreases} 
the thermal conductance, extracted from the data by the data analysis technique \cite{x19-Nature-2018,x24-Nature-2017}.

The reason for the decrease is the subtraction procedure \cite{x19-Nature-2018,x24-Nature-2017}. The thermal conductance of an arm is obtained from the equation

\begin{equation}
\label{dima-new-app-B-1}
K_{\rm arm}=\frac{K_{(N)}-K_{(b)}}{N-b},
\end{equation}
where $N>b$ and $K_{(n)}$ is the observed thermal conductance of the device with $n$ open arms. The procedure allows eliminating unknown bulk losses from the central Ohmic reservoir 
\cite{x21-Science-2013}. Indeed, such losses are the same for any number of open arms. Bulk losses from closed arms are greater for lower $b$. This increases $K_{(b)}$ and decreases
$K_{\rm arm}$ (\ref{dima-new-app-B-1}). 

Crucially, the above effect is present even for chiral edges, such as the edge of a $7/3$ liquid (Section \ref{subsec:7/3}). Thus, we can use the data, obtained at $\nu=7/3$,
to estimate the role of bulk losses. Their role depends strongly on the maximal temperature $T_m$ of the central Ohmic reservoir. The $T_m\sim 45~{\rm mK}$ data are shown in 
Extended Data Fig. 5 of Ref. \onlinecite{x19-Nature-2018}. The expected thermal conductance for $b$ open arms $K_{(b)}T=3b\kappa_0 T$. The observed $K_{(4)}=12.13\kappa_0$ suggests a small contribution, unrelated to edge physics, of $0.13\kappa_0$. The experimental $K_{(3)}=9.26\kappa_0$ and $K_{(2)}=6.36\kappa_0$ are consistent with an additional contribution $\sim 0.1\kappa_0$ per closed edge. Importantly, such corrections become considerably smaller at lower $T_m\sim 30~{\rm mK}$. This suggests that bulk losses are unimportant for such values of $T_m$. For this reason we use the data, obtained at $T_m\sim 30~{\rm mK}$ at $\nu=8/3$ (Fig. 3 in Ref. \onlinecite{x19-Nature-2018}) and ignore bulk losses.

\section{Phenomenological theory of the equilibration at $\nu=8/3$} 
\label{app:pheno-model}

\begin{figure}[htb]
\includegraphics[width=3.0 in]{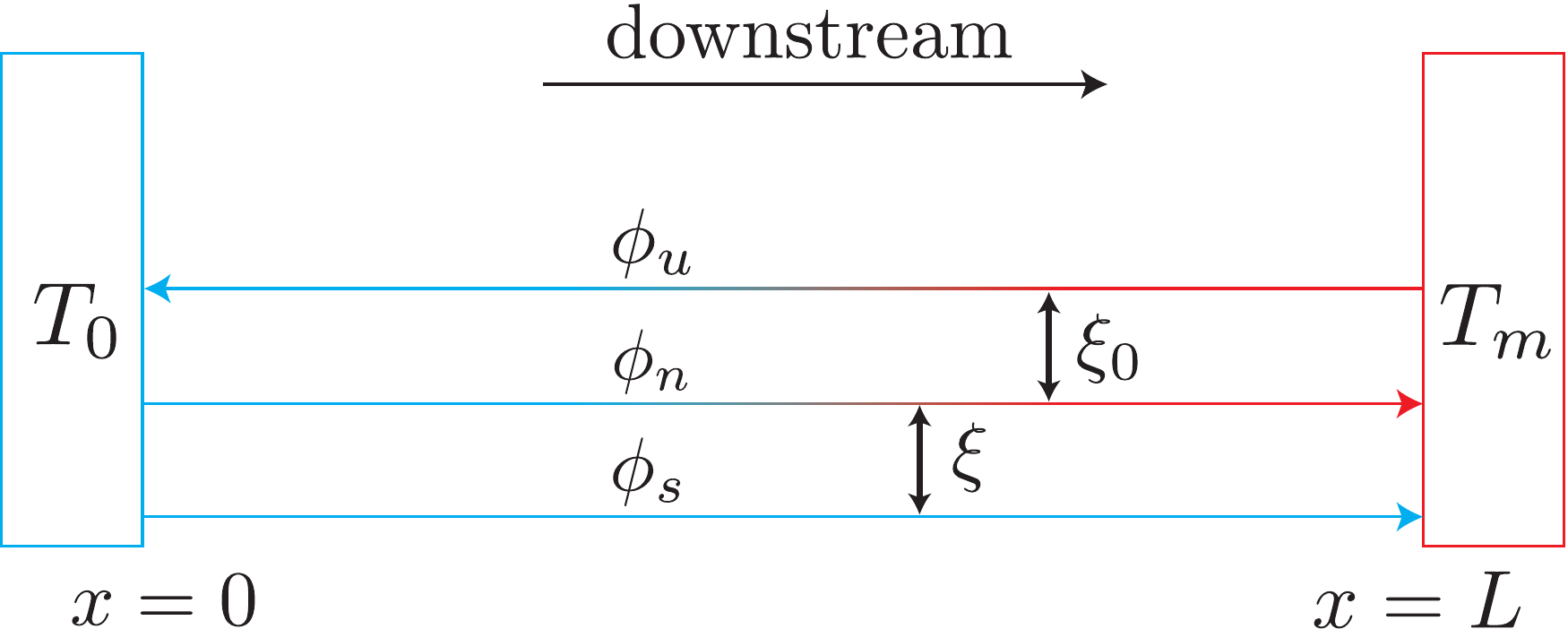}
\caption{(color online) An edge out of equilibrium.}
\label{fig:app1}
\end{figure}

We address the role of the random interaction between $\phi_s$ and $\phi_n$. We do not include any interaction of $\phi_s$ and the upstream mode in the model below. Consider an edge of length $L$, connecting two terminals of the temperatures $T_0$ and $T_m$ (Fig.~\ref{fig:app1}). Neglect energy exchange with the charged mode $\phi_c$. Let the temperature of the upstream mode $T_u(x)$ satisfy $T_u(L)=T_m$. Let the temperature $T_s$ of the spin mode  and the temperature $T_d$ of the downstream mode $\phi_n$ satisfy $T_s(0)=T_d(0)=T_0$. In the spirit of Ref. \onlinecite{x24-Nature-2017}, we write phenomenological equations for the energy balance:
\begin{align}
\label{A1}
\partial_x T_u
&=\frac{T_u-T_d}{\xi_0};
\\
\label{A2}
\partial_x T_d
&=\frac{T_u-T_d}{\xi_0}+\frac{T_s-T_d}{\xi};
\\
\label{A3}
\partial_x T_s
&=\frac{T_d-T_s}{\xi}
\end{align}
with the equilibration lengths $\xi$ and $\xi_0$. Identical equations can be written on the opposite edge of the quantum Hall bar, but $T_0\leftrightarrow T_m$ in the boundary conditions.
The solution of the equations is straightforward. After adding a quantum of heat conductance due to the mode $\phi_c$, we find
\begin{equation}
\label{A4}
K/\kappa_0=2-\frac{2}{1+\frac{rr_0^2}{2\sqrt{r(r+r_0)}}\left[\frac{\exp(L\gamma_-)}{\gamma_-^2}-\frac{\exp(L\gamma_+)}{\gamma_+^2}\right]},
\end{equation}
where $r=1/\xi$, $r_0=1/\xi_0$, and $\gamma_\pm=-r\pm\sqrt{r(r+r_0)}$. With the measured $K/\kappa_0=2.19$, the above equation shows that $\xi_0$ changes between $\approx 0.1L$ and $\approx 0.3L$ in the whole range of $\xi$ from 0 to infinity. For comparison, Ref. \onlinecite{x24-Nature-2017} estimates $\xi_0\sim 0.2L$ at $\nu=2/3$. We conclude that any strength of the random interaction, which determines $\xi$,
is consistent with the data.

\section{Corrections to thermal conductance} 
\label{app:correction}

Consider a group of upstream modes of the total thermal conductance $K_u T$ and a group of downstream modes of the total thermal conductance $K_d T$. We know that in thermal equilibrium the combined thermal conductance of up- and down-stream modes is $|K_u-K_d|T$. The phenomenological theory \cite{x24-Nature-2017} shows that the combined thermal conductance increases away from thermal equilibrium. The positive correction $\Delta K_2 T$ is small if the up- and down-stream modes are close to equilibrium. This happens if the equilibration length
$l_e\ll L$, where $L$ is the edge length \cite{x24-Nature-2017}.

In the opposite limit of $l_e\rightarrow \infty$, the thermal conductance approaches $KT=(K_u+K_d)T$. A large finite $l_e$ results in a small negate correction to the above expression:

\begin{equation}
\label{dima-new-D1}
KT=(K_u+K_d)T-\Delta K_1 T.
\end{equation}
At a fixed temperature difference between the central Ohmic reservoir and a terminal (Fig. \ref{fig:app-dima-1}), the thermal conductance $KT$ is proportional to the thermal current between the reservoir and the terminal. This can be used to estimate $\Delta K_1$. Indeed, thermal exchange between the up- and down-stream modes backscatters a fraction $J_b$ of the thermal current. According to the definition of the equilibration length, $J_b\sim L/l_e$. It follows that

\begin{equation}
\label{dima-new-D2}
\Delta K_1\sim \frac{L}{l_e}.
\end{equation}

$\Delta K_2$ and $-\Delta K_1$ are not the only possible corrections to $K_{5/2}$. Two of the eight edges in the setup \cite{x19-Nature-2018} contain gate-defined regions (Extended Data Fig. 1 in Ref. \onlinecite{x24-Nature-2017}). The gates screen the interaction of the integer and fractional channels. Hence, it is possible that the integer and fractional channels decouple under the gates similarly to the decoupling of channels on the etched edge. At the same time, the transition regions between the etched and gated edges may contribute to the equilibration of all pairs of channels. Note that the hearts of the devices are different in
 Refs. \cite{x24-Nature-2017} and \cite{x19-Nature-2018} so that six of the edges do not contain 
gate-defined regions in the experiment \cite{x19-Nature-2018}. Another subtlety involves possible lack of thermal equilibrium in the central 
Ohmic contact (Appendix \ref{app:RC}).

\section{Possible lack of equilibrium in the central Ohmic contact} 
\label{app:RC}

The $RC$ time $\tau_{RC}=RC$ is the total time that charge spends in a system. According to the uncertainty relations, the uncertainty of the energy change over such time period $\Delta E\sim \hbar/2\tau_{RC}$. This means that the degree of freedom, associated with the total charge, cannot equilibrate with the rest of the system, if 

\begin{equation}
\label{C1}
k_B T\ll k_B T^*= \hbar/2\tau_{RC}.
\end{equation}
This mechanism reduces \cite{theory-RC,exp-RC} the thermal conductance of the setup \cite{x19-Nature-2018,x24-Nature-2017}  by one quantum $\kappa_0T$ at low $T$. 
This is the reduction of the total thermal conductance $NKT$ of an $N$-arm system, Fig. \ref{fig:app-dima-1}. The experiment uses $N=4$ arms and hence the expected reduction of the observed $K$ is $0.25\kappa_0$. Such reduction was not seen at any filling factor, investigated in Refs. \onlinecite{x19-Nature-2018,x24-Nature-2017}. This may appear puzzling. Indeed, the relevant time is the time the charge spends in the central Ohmic contact. One can estimate its capacitance from its known size. It appears that Eq. (\ref{C1}) is satisfied. A possible explanation of this paradox \cite{moty-privite} involves a large contribution of the edge channels to the total capacitance.

Note that the subtraction procedure, employed in Refs. \onlinecite{x19-Nature-2018,x24-Nature-2017}, suppresses the reduction of $K$ due to the $RC$-time constraint. Indeed, $KT$ is defined as one half of the difference of the total thermal conductances $NKT$ in the 4- and 2-arm configurations: $K_{\rm subtraction}T=[K_{(4)}-K_{(2)}]/2$. At a sufficiently low temperature, both $K_{(4)}$ and $K_{(2)}$ are reduced by one quantum $\kappa_0T$. Hence, the reduction effect drops out from their difference. A small reduction of $K_{\rm subtraction}$ is possible in the crossover region between the high- and low-temperature regimes.

\section{Possible breakdown of bulk-edge correspondence}
\label{app:bulk-edge}

The bulk of the experimental evidence in favor of the PH-Pfaffian order comes from edge probes. In the absence of bulk-edge correspondence \cite{wen-book}, such probes shed no light on the bulk physics. Can bulk-edge correspondence break down? We propose a scenario, in which the PH-Pfaffian topological order exists along the edges of the sample \cite{footnote6}. The anti-Pfaffian (or some other) order exists in the bulk of the sample. The gapless interface between the PH-Pfaffian and anti-Pfaffian phases must be far from the edge and the Ohmic contacts. We are not aware of a mechanism behind such a scenario.



\begin{thebibliography}{99}

\bibitem{jain-book}
J. K. Jain, {\it Composite Fermions} (Cambridge University Press, Cambridge, 2007).

\bibitem{x41-RG}
N. Read and D. Green,
{Phys. Rev. B \textbf{61}, 10267 (2000)}.

\bibitem{2-MR} G. Moore and N. Read, Nucl. Phys. B $\mathbf{360}$, 362 (1991).

\bibitem{foot0} For a review of proposed orders see G. Yang and D. E. Feldman, Phys. Rev. B {\bf 88}, 085317 (2013) and Ref. \onlinecite{x19-Nature-2018}.

\bibitem{HR-1988}
F. D. M. Haldane and E. H. Rezayi,
Phys. Rev. Lett. {\bf 60}, 956 (1988); {\it ibid.} 1886 (1988).

\bibitem{myg} 
A. H. MacDonald, D. Yoshioka, and S. M. Girvin,
Phys. Rev. B {\bf 39}, 8044(R) (1989).

\bibitem{x1-Morf}
R. H. Morf, 
{Phys. Rev. Lett. \textbf{80}, 1505 (1998)}.

\bibitem{new-num} 
K. Pakrouski, M. R. Peterson, T. Jolicoeur, V. W. Scarola, C. Nayak, and M. Troyer,
Phys. Rev. X {\bf 5}, 021004 (2015).

\bibitem{Rezayi-2017}
E. H. Rezayi,
Phys. Rev. Lett. {\bf 119}, 026801 (2017) and references therein.

\bibitem{x3-LRH}
M. Levin, B. I. Halperin, and B. Rosenow,
{Phys. Rev. Lett. \textbf{99}, 236806 (2007)}.

\bibitem{x4-LRNF}
S.-S. Lee, S. Ryu, C. Nayak, and M. P. A. Fisher,
{Phys. Rev. Lett. \textbf{99}, 236807 (2007)}.

\bibitem{BoYang}
B. Yang, Phys. Rev. B \textbf{98}, 201101(R) (2018).

\bibitem{x12-ZF}
P. T. Zucker and D. E. Feldman,
{Phys. Rev. Lett. \textbf{117}, 096802 (2016)}.

\bibitem{x18-Son} 
D. T. Son,
{Phys. Rev. X \textbf{5}, 031027 (2015)}.

\bibitem{x18a-Ashwin}
L. Fidkowski, X. Chen, and A. Vishwanath, Phys. Rev. X {\bf 3}, 041016 (2013).

\bibitem{x18b-Bonderson}
P. Bonderson, C. Nayak, and X.-L. Qi, J. Stat. Mech. {\bf 2013}, P09016.

\bibitem{disorder-num}
Very recently, disorder was addressed in W. Zhu and D. N. Sheng, arXiv:1809.04776.

\bibitem{x19-Nature-2018}
M. Banerjee, M. Heiblum, V. Umansky, D. E. Feldman, Y. Oreg, and A. Stern, Nature {\bf 559}, 205 (2018).

\bibitem{x22-KF-eq}
C. L. Kane and M. P. A. Fisher,
Phys. Rev. B {\bf 55}, 15832 (1997).

\bibitem{Capelli}
A. Cappelli, M. Huerta, and G. R. Zemba, Nucl. Phys. B {\bf 636}, 568 (2002).

\bibitem{x21-Science-2013}
S. Jezouin, F. D. Parmentier, A. Anthore, U. Gennser, A. Cavanna, Y. Jin, and F. Pierre,
Science  {\bf 342}, 601 (2013).


\bibitem{x24-Nature-2017}
M. Banerjee, M. Heiblum, A. Rosenblatt, Y. Oreg, D. E. Feldman, A. Stern, and V. Umansky, Nature (London) {\bf 545}, 75 (2017).

\bibitem{x43-d1}
D. F. Mross, Y. Oreg, A. Stern, G. Margalit, and M. Heiblum, Phys. Rev. Lett. {\bf 121}, 026801 (2018).

\bibitem{x44-d2}
C. Wang, A. Vishwanath, B. I. Halperin, Phys. Rev. B {\bf 98}, 045112 (2018).

\bibitem{x45-d3}
B. Lian and J. Wang, Phys. Rev. B {\bf 97}, 165124 (2018).

\bibitem{Milovanovic}
L. Antoni\'c, J. Vu\v{c}i\v{c}evi\'c, and M. V. Milovanovi\'c,
Phys. Rev. B {\bf 98}, 115107 (2018).

\bibitem{x25-Simon}
S. H. Simon, Phys. Rev. B {\bf 97}, 121406(R) (2018).

\bibitem{comment}
D. E. Feldman, Phys. Rev. B {\bf 98}, 167401 (2018).

\bibitem{8-3-2011}
M. Dolev, Y. Gross, R. Sabo, I. Gurman, M. Heiblum, V. Umansky, and D. Mahalu,
Phys. Rev. Lett. {\bf 107}, 036805 (2011).

\bibitem{Tiemann-NMR}
L. Tiemann, G. Gamez, N. Kumada, and K. Muraki, Science {\bf 335}, 828 (2012).

\bibitem{Pan-spin}
W. Pan, K. W. Baldwin, K. W. West, L. N. Pfeiffer, and D. C. Tsui, Phys. Rev. Lett. {\bf 108}, 216804 (2012).


\bibitem{wen-book}
X.-G. Wen, {\it Quantum Field Theory of Many-Body Systems: From the Origin of Sound to an Origin of Light and Electrons} (Oxford University Press, 2004).

\bibitem{Stern-NMR}
M. Stern, B. A. Piot, Y. Vardi, V. Umansky, P. Plochocka, D. K. Maude, and I. Bar-Joseph, Phys. Rev. Lett. {\bf 108}, 066810 (2012).

\bibitem{Stern-optical}
M. Stern, P. Plochocka, V. Umansky, D. K. Maude, M. Potemski, and I. Bar-Joseph, Phys. Rev. Lett. {\bf 105}, 096801 (2010).

\bibitem{Rhone-2011}
T. D. Rhone, J. Yan, Y. Gallais, A. Pinczuk, L. Pfeiffer, and K. West,
Phys. Rev. Lett. {\bf 106}, 196805 (2011).

\bibitem{Wurstbauer2013}
U. Wurstbauer, K. W. West, L. N. Pfeiffer, and A. Pinczuk, Phys. Rev. Lett. {\bf 110}, 026801 (2013).



\bibitem{x40-composite}
R. L. Willett, K. W. West, and L. N. Pfeiffer,
Phys. Rev. Lett. {\bf 88}, 066801 (2002).

\bibitem{xnew-composite}
Md. Shafayat Hossain, M. K. Ma, M. A. Mueed, L. N. Pfeiffer, K. W. West, K. W. Baldwin, and M. Shayegan,
Phys. Rev. Lett. {\bf 120}, 256601 (2018).

\bibitem{num-spin-1}
A. E. Feiguin, E. Rezayi, Kun Yang, C. Nayak, and S. Das Sarma,
Phys. Rev. B {\bf 79}, 115322 (2009).

\bibitem{num-spin-2}
J. Biddle, M. R. Peterson, and S. Das Sarma,
Phys. Rev. B {\bf 87}, 235134 (2013).

\bibitem{num-spin-3}
I. Dimov, B. I. Halperin, and C. Nayak,
Phys. Rev. Lett. {\bf 100}, 126804 (2008).\bibitem{Pan-transition}
W. Pan, A. Serafin, J. S. Xia, L. Yin, N. S. Sullivan, K. W. Baldwin, K. W. West, L. N. Pfeiffer, and D. C. Tsui, Phys. Rev. B {\bf 89}, 241302(R) (2014).

\bibitem{Samkharadze2017}
N. Samkharadze, D. Ro, L. N. Pfeiffer, K. W. West, and G. A. Cs\'{a}thy, Phys. Rev. B {\bf 96}, 085105 (2017).

\bibitem{Falson-ZnO}
J. Falson, D. Maryenko, B. Friess, D. Zhang, Y. Kozuka, A. Tsukazaki, J. H. Smet, and M. Kawasaki,
Nature Phys. {\bf 11}, 347 (2015).

\bibitem{8-5-polarized}
R. R. Du, A. S. Yeh, H. L. Stormer, D. C. Tsui, L. N. Pfeiffer, and K. W. West,
Phys. Rev. Lett. {\bf 75}, 3926 (1995).

\bibitem{so-2018}
A. Aharon, Y. Oreg, and A. Stern, arXiv:1805.09229.

\bibitem{eq1-spin}
G. M\"uller, D. Weiss, A. V. Khaetskii, K. von Klitzing, S. Koch, H. Nickel, W. Schlapp, and R. L\"osch, Phys. Rev. B {\bf 45}, 3932 (1992).

\bibitem{KFP}
C. L. Kane, M. P. A. Fisher, and J. Polchinski, Phys. Rev. Lett. {\bf 72}, 4129 (1994).

\bibitem{KF-randomness}
C. L. Kane and M. P. A. Fisher,
Phys. Rev. B {\bf 51}, 13449 (1995).

\bibitem{CCG}
D. B. Chklovskii, B. I. Shklovskii, and L. I. Glazman,
Phys. Rev. B {\bf 46}, 4026 (1992).

\bibitem{GH}
B. Y. Gelfand and B. I. Halperin, Phys. Rev. B {\bf 49}, 1862 (1994).

\bibitem{eq2-hirai}
H. Hirai, S. Komiyama, S. Fukatsu, T. Osada, Y. Shiraki, and H. Toyoshima,
Phys. Rev. B {\bf 52}, 11159 (1995).

\bibitem{eq3-deviatov}
A. W\"urtz, R. Wildfeuer, A. Lorke, E. V. Deviatov, and V. T. Dolgopolov,
Phys. Rev. B {\bf 65}, 075303 (2002).

\bibitem{edge-width-1}
J. B. Miller, I. P. Radu, D. M. Zumb\"uhl, E. M. Levenson-Falk, M. A. Kastner, C. M. Marcus, L. N. Pfeiffer, and K. W. West,
Nature Phys. {\bf 3}, 561 (2007).

\bibitem{edge-width-2}
I. P. Radu, J. B. Miller, C. M. Marcus, M. A. Kastner, L. N. Pfeiffer, and K. W. West, Science {\bf 320}, 899 (2008).

\bibitem{baer} 
S. Baer, C. R\"ossler, T. Ihn, K. Ensslin, C. Reichl, and W. Wegscheider, Phys. Rev. B {\bf 90}, 075403 (2014).

\bibitem{kumada-v}
N. Kumada, H. Kamata, and T. Fujisawa, Phys. Rev. B {\bf 84}, 045314 (2011).

\bibitem{feve}
E. Bocquillon, V. Freulon, J.-M. Berroir, P. Degiovanni, B. Pla\c{c}ais, A. Cavanna, Y. Jin, and G. F\'eve, Nat. Commun. {\bf 4}, 1839 (2013).

\bibitem{kun-yang-2-3}
Z.-X. Hu, H. Chen, K. Yang, E. H. Rezayi, and X. Wan, Phys. Rev. B {\bf 78}, 235315 (2008).

\bibitem{hashisaka}
M. Hashisaka, N. Hiyama, T. Akiho, K. Muraki, and T. Fujisawa, Nature Phys. {\bf 13}, 559 (2017).

\bibitem{charge-eq}  
F. Lafont, A. Rosenblatt, M. Heiblum, and V. Umansky, arXiv:1808.05898.

\bibitem{bosonization}
J. von Delft and H. Schoeller, Ann. Phys. {\bf 7}, 225 (1998).

\bibitem{spin-dima-new}
R. J. Nicholas, R. J. Haug, K. v. Klitzing, and G. Weimann,
Phys. Rev. B {\bf 37}, 1294 (1988).

 

\bibitem{xilin1}
X. Lin, C. Dillard, M. A. Kastner, L. N. Pfeiffer, and K. W. West, Phys. Rev. B {\bf 85}, 165321 (2012).

\bibitem{xilin2}
H. Fu, P. Wang, P. Shan, L. Xiong, L. N. Pfeiffer, K. West, M. A. Kastner, and X. Lin, PNAS {\bf 113}, 12386 (2016).

\bibitem{theory-RC}
A. O. Slobodeniuk, I. P. Levkivskyi, and E. V. Sukhorukov,
Phys. Rev. B {\bf 88}, 165307 (2013).

\bibitem{exp-RC}
E. Sivre, A. Anthore, F. D. Parmentier, A. Cavanna, U. Gennser, A. Ouerghi, Y. Jin, and F. Pierre,
Nature Phys. {\bf 14}, 145 (2018).

\bibitem{moty-privite} M. Heiblum, private communication.

\bibitem{footnote6} 
Ref. \onlinecite{x44-d2} considers a scenario with different orders in the bulk and near the edge.

\end{thebibliography}
\end{document}